\documentclass[usenatbib]{mn2e}

\usepackage{epsfig}
\usepackage{subfigure}
\usepackage{amssymb}
\usepackage{amsmath}

\newcommand{\bn}{\begin{enumerate}}
\newcommand{\en}{\end{enumerate}}
\newcommand{\bi}{\begin{itemize}}
\newcommand{\ei}{\end{itemize}}

\newcommand{\Msun}{M_\odot}

\newcommand{\himpc}{h^{-1} {\rm Mpc}}
\newcommand{\hikpc}{h^{-1} {\rm kpc}}

\newcommand{\Muv}{M_{\rm uv}}

\newcommand{\apj}{ApJ}
\newcommand{\aap}{A\&A}
\newcommand{\apjl}{ApJL}
\newcommand{\apjs}{ApJS}
\newcommand{\mnras}{MNRAS}
\newcommand{\aj}{AJ}
\newcommand{\araa}{ARA\&A}

\title[Steep Faint-end Slopes of Galaxy Mass and Luminosity Functions at $z\geq 6$]{Steep Faint-end Slopes of Galaxy Mass and Luminosity Functions at $z\geq 6$ and the Implications for Reionisation}
\author[Jaacks et al.]{Jason Jaacks$^1$\thanks{Email: jaacksj@physics.unlv.edu}, Jun-Hwan Choi$^2$, Kentaro Nagamine$^1$, Robert Thompson$^1$, Saju Varghese$^1$ \vspace{0.3cm}\\
$^1$ Department of Physics \& Astronomy, University of Nevada, Las Vegas, 4505 S. Maryland Pkwy, Las Vegas, NV, 89154-4002, USA \\
$^2$ Department of Physics \& Astronomy, University of Kentucky, Lexington, KY 40506, USA.
}

\begin{document}

\maketitle
 
\begin{abstract}
We present the results of a numerical study comparing photometric and physical properties of simulated $z=6-9$ galaxies to the observations taken by the WFC3 instrument aboard the Hubble Space Telescope.  Using cosmological hydrodynamical simulations we find good agreement with observations in color-color space at all studied redshifts.  We also find good agreement between observations and our Schechter luminosity function fit in the observable range, $\Muv\leq -18$, provided that a moderate dust extinction effect exists for massive galaxies.  However beyond what currently can be observed, simulations predict a very large number of low-mass galaxies and evolving steep faint-end slopes from $\alpha_L = -2.15$ at $z=6$ to $\alpha_L = -2.64$ at $z=9$, with a dependence of $|\alpha_L| \propto (1+z)^{0.59}$.  During the same epoch, the normalization $\phi^{*}$ increases and the characteristic magnitude $\Muv^*$ becomes moderately brighter with decreasing redshift. We find similar trends for galaxy stellar mass function with evolving low-mass end slope from $\alpha_M = - 2.26$ at $z=6$ to $\alpha_M = -2.87$ at $z=9$, with a dependence of $|\alpha_M| \propto (1+z)^{0.65}$.  Together with our recent result on the high escape fraction of ionizing photons for low-mass galaxies, our results suggest that the low-mass galaxies are important contributor of ionizing photons for the reionisation of the Universe at $z\ge 6$. 
\end{abstract}


%%%%%%%%%%%%%%%%%%%%%%%%%%%%%%%%%%%%%%%%%%%%%%%%%%

\section{Introduction}
\label{sec:intro}
When and how galaxies formed throughout the history of the Universe is one of the most fundamental questions of astronomy and astrophysics. As technology improves, astronomers are able to push the redshift frontier of galaxy observation and attempt to obtain a glimpse of galaxy formation processes at high redshift. 
In particular, upgrades to the Hubble Space Telescope (HST) via the addition of the Wide Field Camera 3 (WFC3)  have expanded our view of the Universe to reach redshifts $z\gtrsim 7$, and dozens of new high-redshift candidates have been identified \citep[e.g.,][]{Yan&Windhorst.etal:09, Oesch.etal:09, Ouchi.etal:09, Bouwens.etal:10a,Wilkins.etal:11a}.   For example, \citet{Bouwens.etal:10a} identified $113$ $z\sim 7 - 8$ Lyman-Break candidates by color-color selection with AB magnitudes $\Muv \simeq 26-29$ and faint-end luminosity slopes of $\alpha_L=-1.94\pm0.24$ at $z\sim7$ and $\alpha_L= -2.00\pm0.33$ at $z\sim8$.

While these new observational results are very exciting, candidate galaxies at $z\ge 7$ are very faint and it is still difficult to obtain high-resolution spectra to estimate their physical properties such as stellar masses and star formation rates (SFR). Therefore it would be useful to obtain predictions from theoretical models on their spectro-photometric properties.  In particular, cosmological hydrodynamic simulations based on the standard concordance $\Lambda$ cold dark matter (CDM) model can directly simulate the dynamical evolution of dark matter and baryons from the initial condition that is consistent with the observations of cosmic microwave background. 
With the implementation of radiative heating and cooling, one can also follow the thermal properties of intergalactic medium and collapse of gas clouds into galaxies.  Much work has been done for comparing simulated galaxies at $z\le 6$ to observations \citep[e.g.,][]{Nagamine:04, Nagamine:05a, Nagamine:05b, Night.etal:06,Finlator.etal:06,Robertson.etal:07}, but the comparisons of photometric properties of galaxies at $z\ge7$ between simulations and observations have not been performed adequately yet.

\begin{table*} 
\begin{center}
\begin{tabular}{ccccccc}
\hline
Run  & Box size & $N_{p}$ & $m_{DM}$ & $m_{\rm gas}$ & $\epsilon$ & $z_{\rm end}$ \\ 
& ($\himpc$) & (Dark Matter, Gas)  & ($h^{-1} \Msun$) & ($h^{-1} \Msun$) & ($\hikpc$) \\
\hline
N400L10 & $10.0$ & $2 {\times} 400^{3}$ & $9.37{\times} 10^{5}$ & $1.91 {\times} 10^{5}$ & $1.0$ & $2.75$ \\
N400L34 & $33.75$ & $2 {\times} 400^{3}$ & $3.60 {\times} 10^{7}$ & $7.34 {\times} 10^{6}$ & $3.38$ & $1$ \\
N400L100 & $100.0$ & $2 {\times} 400^{3}$ & $9.37 {\times} 10^{8}$ & $1.91 {\times} 10^{8}$ & $6.45$ & $0$ \\
N600L100 & $100.0$ & $2 {\times} 600^{3}$ & $2.78 {\times} 10^{8}$ & $5.65 {\times} 10^{7}$ & $4.30$ & $0$ \\
\hline
\end{tabular}
\caption{Simulation parameters used in this paper. The parameter $N_p$ is the number of gas and dark matter particles; $m_{\rm DM}$ and $m_{\rm gas}$ are the particle masses of dark matter and gas; $\epsilon$ is the comoving gravitational softening length; $z_{\rm end}$ is the ending redshift of each simulation.  }
\label{tbl:Sim}
\end{center}
\end{table*}

One of the important missing information from theoretical predictions is the precise determinations of evolution of galaxy luminosity function (LF) and stellar mass function (GSMF) at $z\ge 6$.  Some of the earlier numerical work have suggested very steep faint-end slope of $\alpha_L \approx -2$ at $z\approx 6$ \citep{Night.etal:06, Finlator.etal:06}, but its evolution has not been quantified yet at $z\ge 6$ using cosmological hydrodynamic simulations.  This is an important topic, because the number density of early dwarf galaxies has significant implications for the reionisation of the Universe at $z\ge 6$.  It is thought that small dwarf galaxies, that are currently beyond detection, may have played an important role in the reionisation of the Universe \citep[e.g.,][]{Munoz.etal:10, Trenti.etal:10, Yajima.etal:10}. 

In this paper, we compare the results of our cosmological simulations with the recent WFC3 observations. In particular, we will determine the evolution of faint-end (low-mass end) slope $\alpha_L$ ($\alpha_M$) at $z=6-9$ more precisely in our simulations.  The paper is organized as follows:  Section~\ref{sec:simulation} will briefly describe the simulation and setup that was used to obtain our data;  Section~\ref{sec:method} will outline the methods used in processing the photometric data;  Sections~\ref{sec:color-color}-\ref{sec:MFs} will detail assembled data from simulations; Section~\ref{sec:reion} will discuss the implications of these findings to the epoch of reionisation.  Finally in Section~\ref{sec:summary} we will summarise our findings.

%%%%%%%%%%%%%%%%%%%%%%%%%%%%%%%%%%%%%%%%%%%%%%%%%%

\section{Simulations}
\label{sec:simulation}
We use a modified version of the smoothed particle hydrodynamics (SPH) code GADGET-3 \citep[originally described in][]{Springel:05}.
This modified code includes radiative cooling by H, He, and metals \citep{Choi:09}, heating by a uniform UV background of a modified \citet{Haardt:96} spectrum \citep{Katz:96, Dave:99},  an \cite{Eisenstein&Hu:99} initial power spectrum, star formation via the "Pressure model" \citep{Schaye:08, Choi:10a}, supernova feedback, sub-resolution model of multiphase ISM \citep{Springel:03}, and the multicomponent variable velocity (MVV) wind model \citep{Choi:11b}.   We also explore the effect of a new UV background prescription by \citet{Faucher.etal:09} in Section~\ref{sec:h2}.  Our current simulations do not include the AGN feedback.  

We have shown earlier that the metal line cooling enhances star formation across all redshifts by about $10-30$\% \citep{Choi:09}, and that the Pressure SF model suppresses star formation at high-redshift due to a higher threshold density for SF \citep{Choi:10a} with respect to the earlier model by \citet{Springel:03}. \citet{Choi:11b} also showed that the MVV wind model, which is based on the momentum-driven wind, makes the faint-end slope of GSMF slightly shallower compared to the constant velocity galactic wind model of \citet{Springel:03}. One of the points of this paper is to report that, even with these improvements to the supernova feedback models, our simulations continue to predict very steep faint-end slopes. 

It is known that the collisional ionization equilibrium (CIE) table of \citet{Sutherland:93} overestimates the cooling rates compared to the case when the photoionisation by the UVB is taken into account \citep{Efstathiou:92, Wiersma.etal:09}, and \citet{Choi:09} have used this CIE table.  We have compared our LF results in the two runs with and without the metal cooling, and found that there is little difference in the faint-end slope of the two simulations. Therefore this issue is not a problem for the present work. 

For this paper, simulations are setup with either $2\times 400^3$ or $2\times 600^3$ particles for both gas and dark matter. Multiple runs were made with comoving box sizes of $10h^{-1}$Mpc, $34h^{-1}$Mpc and $100h^{-1}$Mpc.  These runs will be referred to as N400L10, N400L34 and N600L100, respectively. Simulation parameters are summarised in Table~\ref{tbl:Sim}.  The adopted cosmological parameters are based on the most recent WMAP data: ${\Omega_{\rm m}}=0.26$, ${\Omega_{\Lambda}}=0.74$, ${\Omega_{\rm b}}=0.044$, $h=0.72$, $\sigma_{8}=0.80$, $n_{s} =0.96$  \citep{Komatsu.etal:10}.

At each time step of the simulation, the gas particles that exceed the SF threshold density ($n_{\rm th}^{\rm SF}=0.6$\,cm$^{-3}$) is allowed to spawn star particles with the probability consistent with the intended SFR. \citet{Choi:10a} have shown that the Pressure SF model with the above $n_{\rm th}^{\rm SF}$ produces favorable results compared to the observed \citet{Kennicutt:98} law, in particular at the low column density end.   Galaxies are then identified by a group-finder algorithm using a simplified variant of the SUBFIND algorithm \citep{Springel:01}.  Collections of star and gas particles are grouped based on the baryonic density field.  Properties such as stellar mass, SFR, metallicity and position, among others are recorded in galaxy property files.  A more detailed description of the group-finder process can be found in \citet{Nagamine:04}.

%%==================================================================

\section{Method}
\label{sec:method}

%%Fig. 1
\begin{figure*}
\begin{center}
\includegraphics[width=1.0\columnwidth,angle=0] {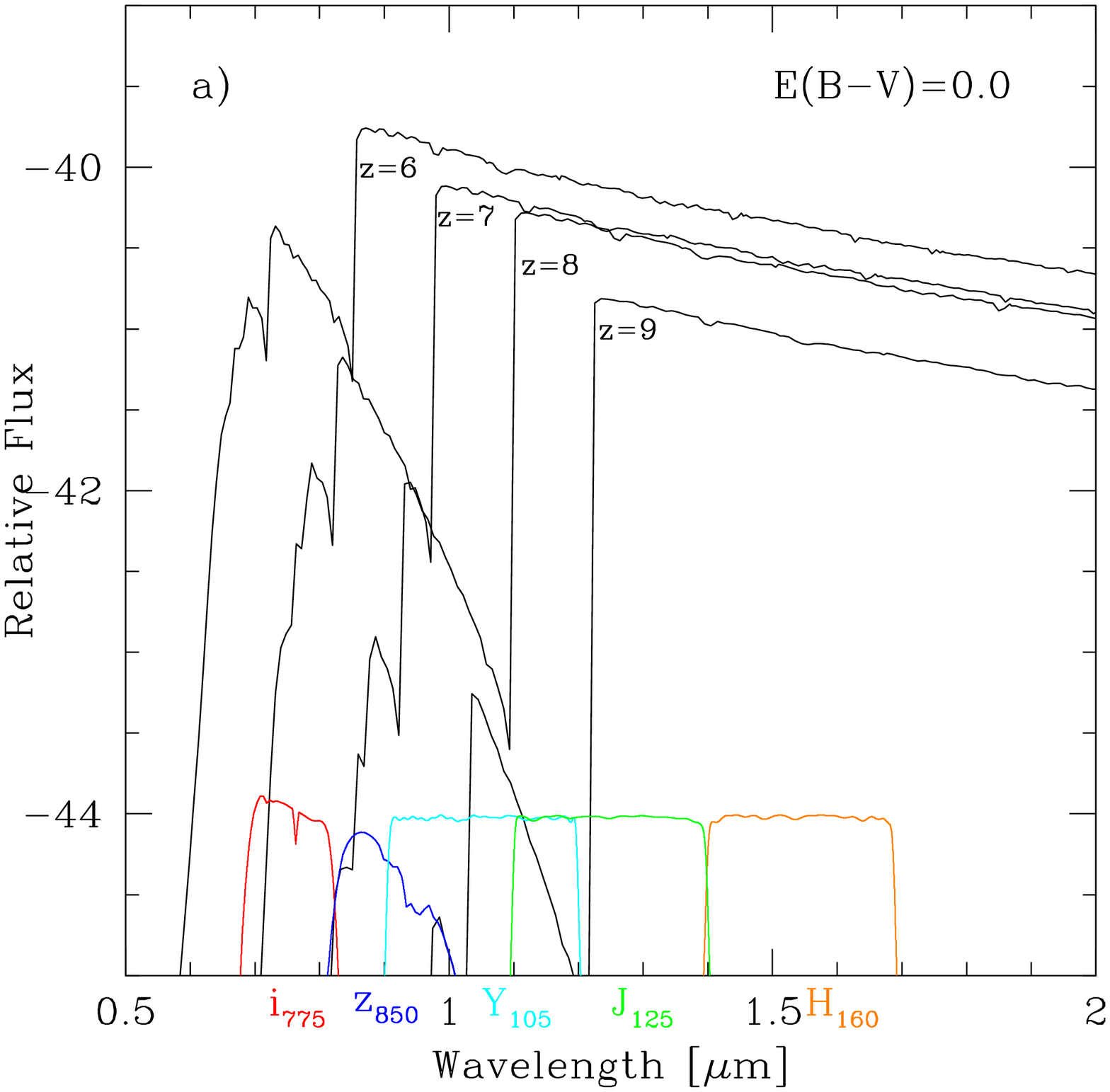}
\includegraphics[width=1.0\columnwidth,angle=0] {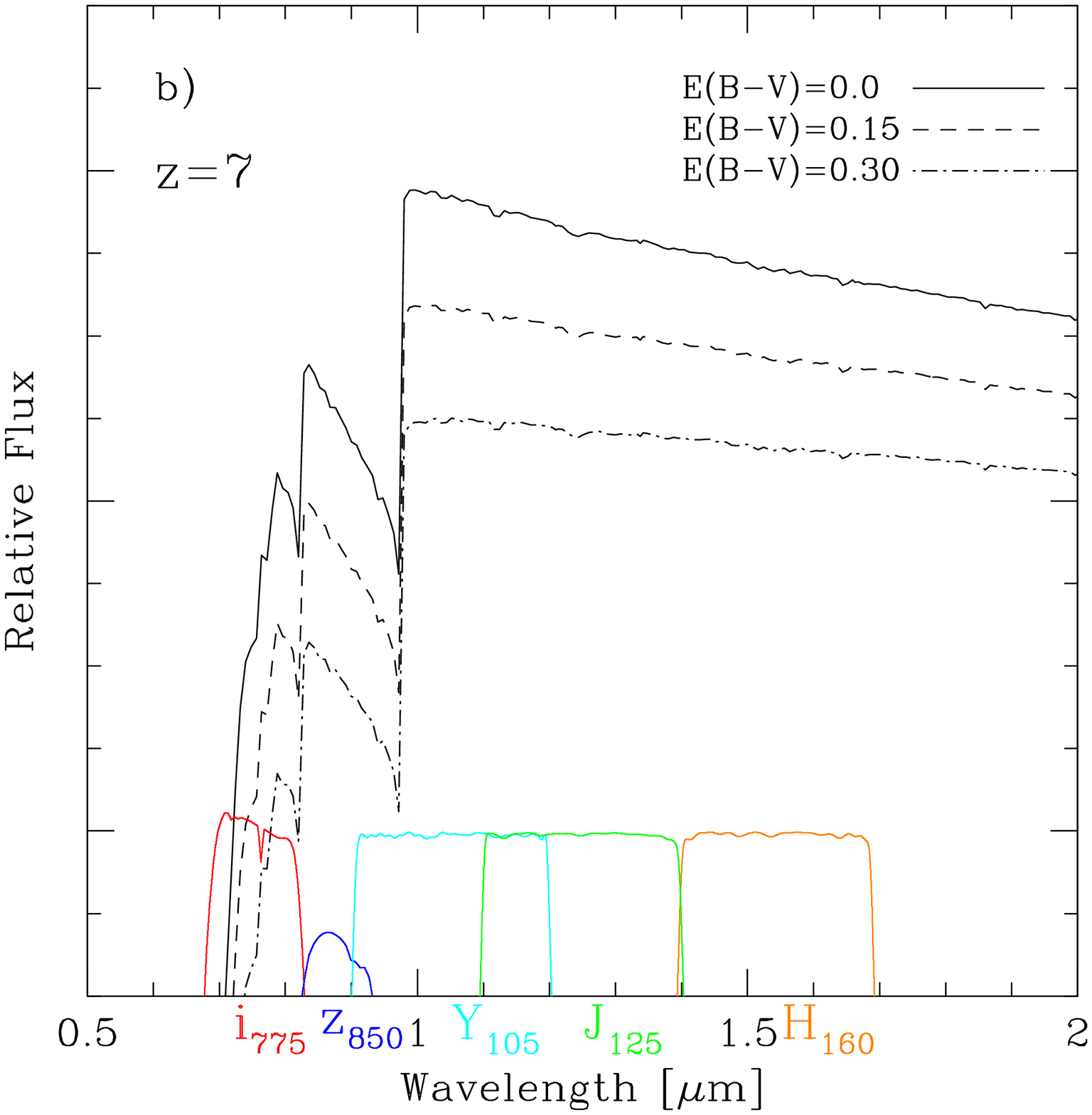}
\caption{
{\it Panel (a)} shows the observed-frame relative flux $f_{\nu}$ of a simulated galaxy in the N400L34 run, redshifted from $z=6$ to $z=9$ with no extinction.  The wedge-like feature on the blue-side is caused by the \citet{Madau:95} prescription of IGM absorption. 
{\it Panel (b)} shows the effect of \citet{Calzetti:97} extinction on the simulated galaxy spectrum at $z=7$ for $E(B-V)=0.0, 0.15,$ \& 0.30. 
}
\label{fig:fnu}
\end{center}
\end{figure*}

Once we identify the individual galaxies, we calculate the spectral energy distribution (SED) of each galaxy by applying the 2007 version of \citet{BC03}  stellar population synthesis model that includes the contribution of TP-AGB phase.  We have repeated the calculation using the 2003 version of \citet{BC03}, and confirmed that the results hardly change.  This is expected because we are focusing our analyses on the rest-frame UV wavelengths range at $z>6$, and the effect of TP-AGB appears only in the infra-red range. In the present work, nebular emission lines are not included in the SEDs, but our results are not affected because this work is mainly concerned about the comparisons in the rest-frame UV, where the optical emission lines do not come into the filters.  We will investigate the impact of nebular emission lines on the estimates of stellar masses of high-redshift galaxies in the subsequent work. 

After we calculate the intrinsic SEDs, we then account for the dust extinction effect using the \citet{Calzetti:97} extinction law.  We adopt uniform constant values of $E(B-V)= 0.0, 0.075, 0.10, 0.15$ \& 0.30.  
The above extinction values are selected based on the varied data found in recent publications.  \citet{Bouwens.etal:09} claimed a trend of decreasing extinction approaching $E(B-V)=0$ at $z\simeq 6$ based on the UV continuum slope.  However  \citet{Schaerer.deBarros:10} performed SED fits including the effect of nebular emission lines, and showed that dust attenuation of up to $E(B-V)\sim 0.32$ ($A_V\sim 1$ for $R_V=3.1$) can be found for galaxies at $z=6-8$, and that $E(B-V)$ might be greater for higher mass galaxies. Our final choice for $E(B-V)$ at each redshift is based on a visual match of the LF to the observed data range, staying within the above mentioned range.  In Section \ref{sec:LFs} we also examine the effect of variable extinction models.

The final step to achieving a spectrum as it would appear to HST, is to factor in the absorption by the intergalactic medium (IGM).  To do this we apply the model of  \citet{Madau:95}. Figure~\ref{fig:fnu}a shows the relative flux $f_{\nu}$ of a representative galaxy redshifted to $z=7, 8$ \& 9.  The HST filter functions used by the observers are also shown on the X-axis: $i_{775}$, $z_{850}$, $Y_{105}$, $J_{125}$, $H_{160}$ centered at $0.775$, $0.850$, $1.05$, $1.25$, $1.60$  $\micron$, respectively, are some of the HST filters strategically positioned to identify the Lyman break of high redshift galaxies.
Figure~\ref{fig:fnu}b shows the effect of \citet{Calzetti:97} extinction on the simulated galaxy spectrum at $z=7$ for $E(B-V)=0.0, 0.15,$ \& 0.30.

The observer-frame spectrum is then used to calculate the apparent magnitude of each galaxy for each HST filter, as well as the rest-frame UV magnitude centered at 1700\AA.  This process gives us a rest-frame magnitude in each filter that can then be assembled into color-color plots and luminosity functions (LFs) for direct comparison to those produced by observers.

%====================================================================

%\section{Results}
%\label{sec:results}
\section{Color-Color Diagrams}
\label{sec:color-color}
\begin{figure*}
\includegraphics[width=0.65\columnwidth,angle=0] {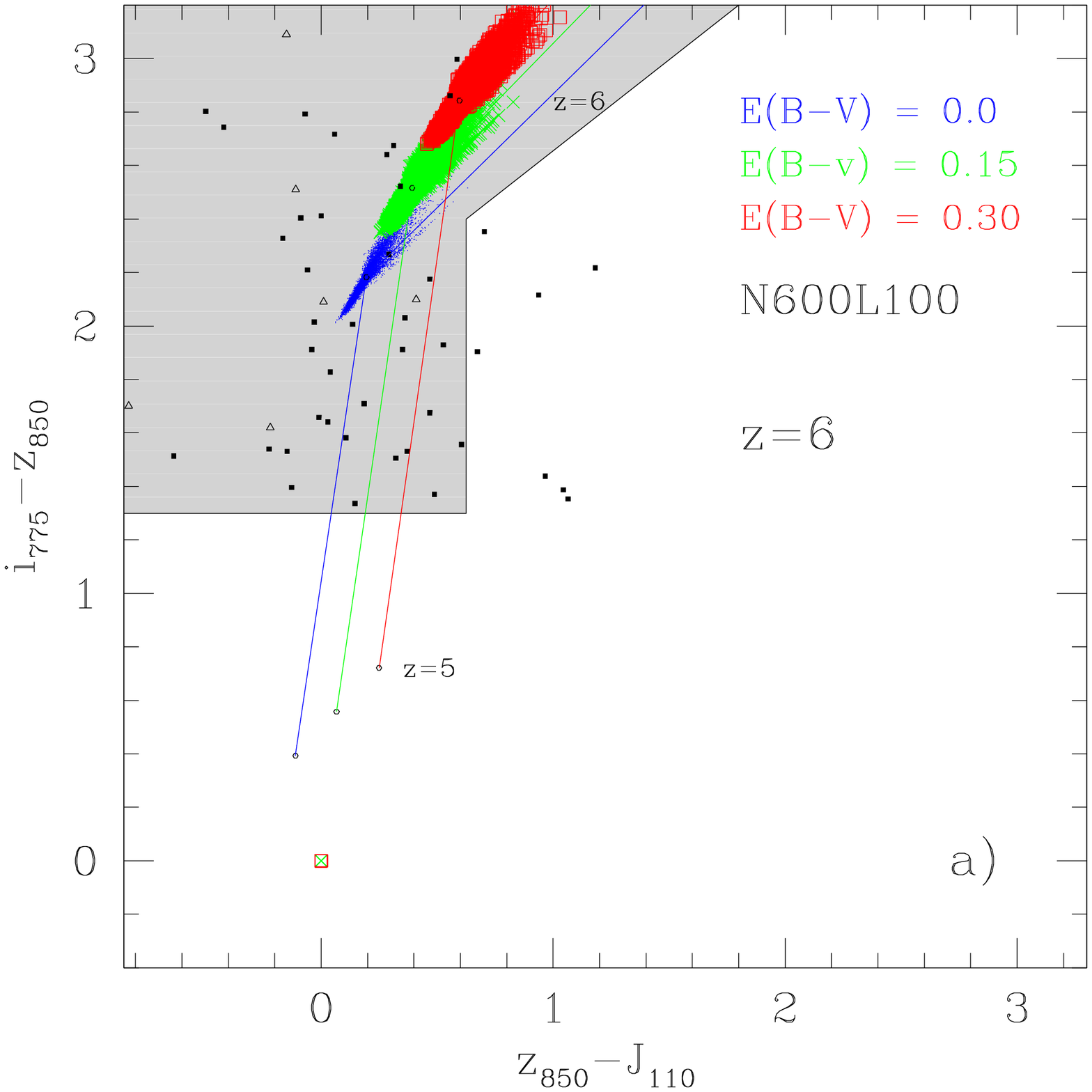}
\includegraphics[width=0.65\columnwidth,angle=0] {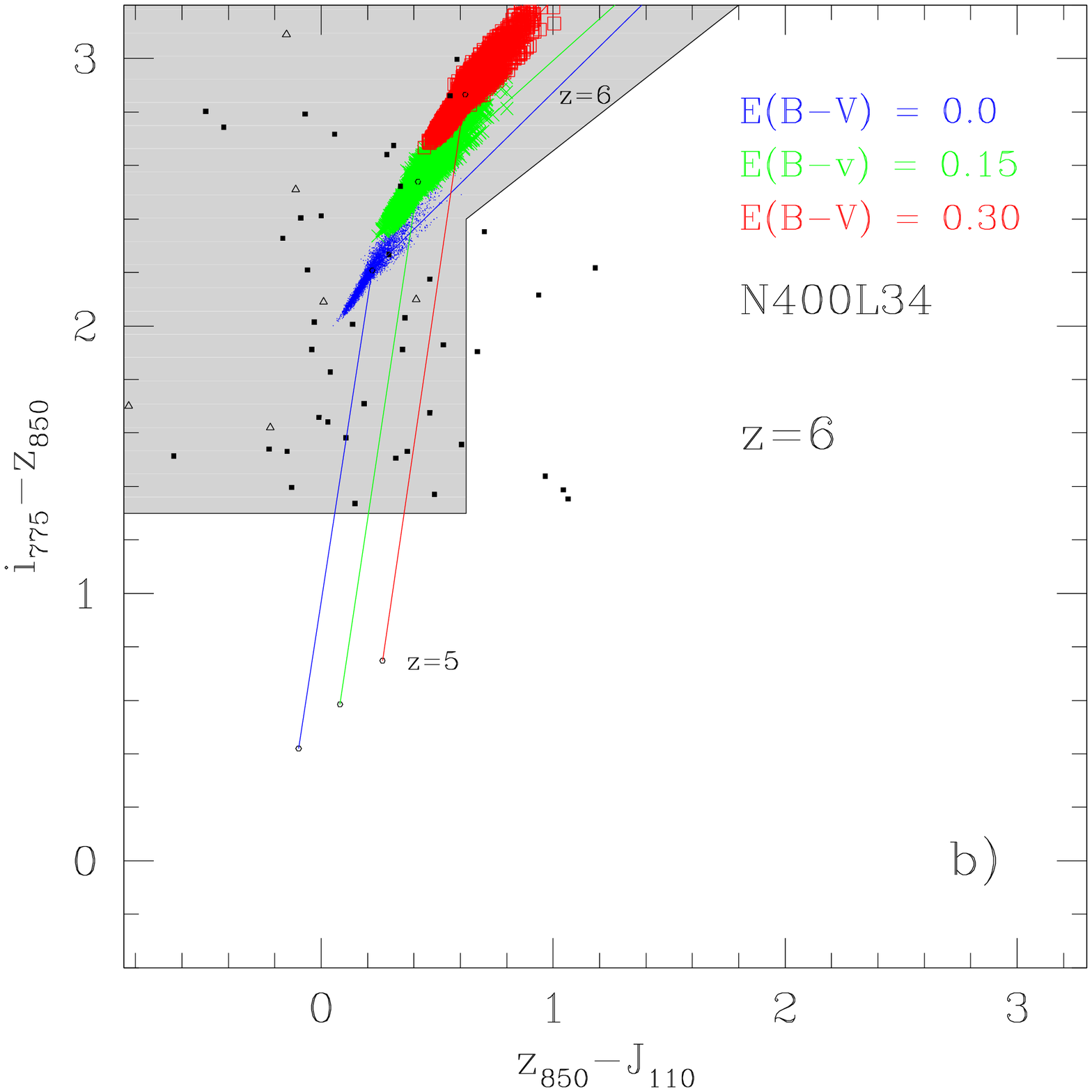}
\includegraphics[width=0.65\columnwidth,angle=0] {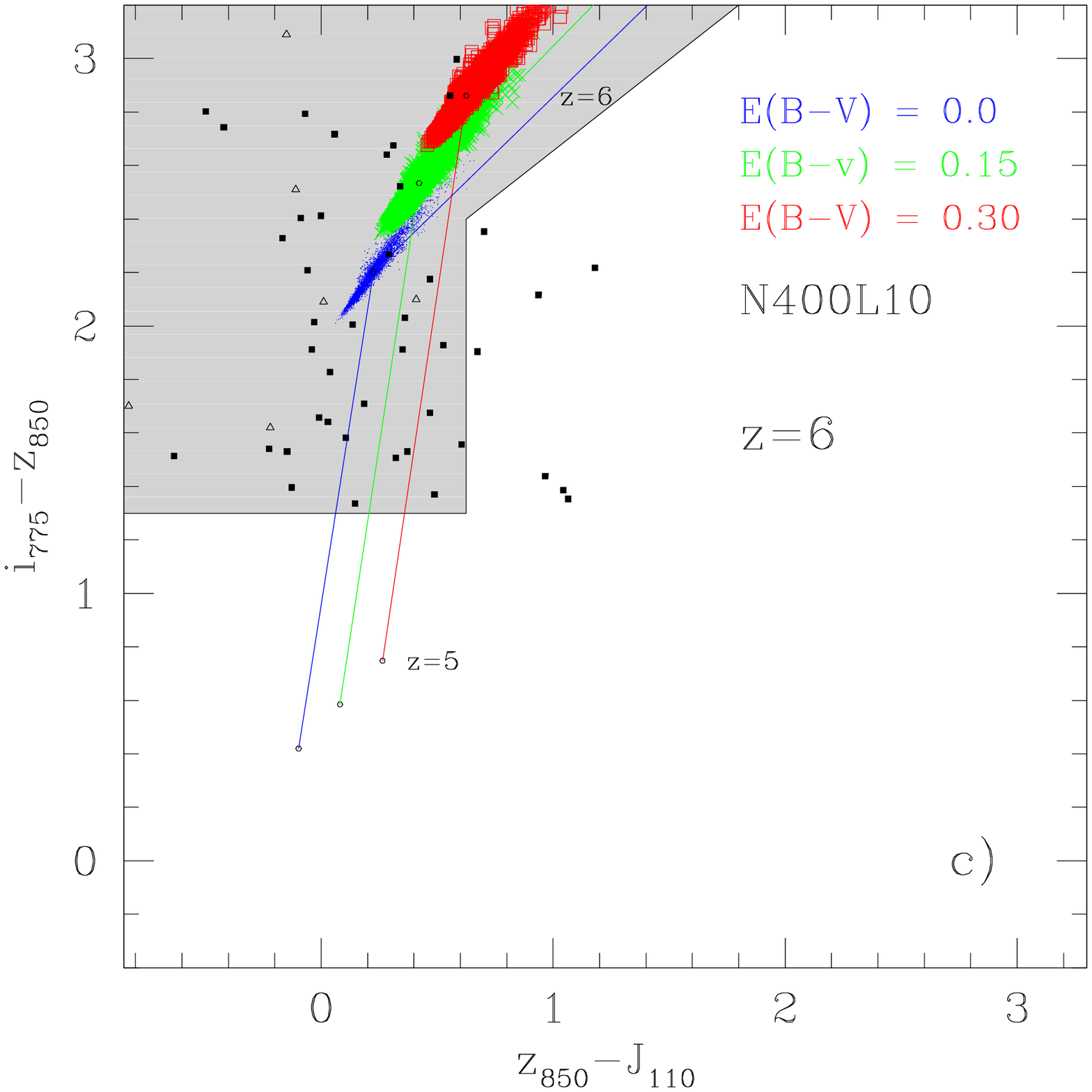}
\caption{
Color-color diagram comparing observed $z\sim 6$ candidates with our simulations of differing box sizes.  The red, green and blue data points represent simulated galaxies with $E(B-V) = 0.0, 0.15, \& 0.30$, respectively.  For each extinction, the color track of a representative galaxy is shown with a solid line from $z=5$ through $z=7$ (out of plot range).  The grey shaded area, defined by \citep{Bouwens.etal:06}, indicates the expected position of $z\sim 6$ objects.  Black filled squares \citep{Bouwens.etal:06} and open triangles \citep{Yan&Windhorst:04} indicate observed $z\sim 6$ candidates. 
}
\label{fig:ccz6}
\end{figure*}

\begin{figure*}
\includegraphics[width=0.65\columnwidth,angle=0] {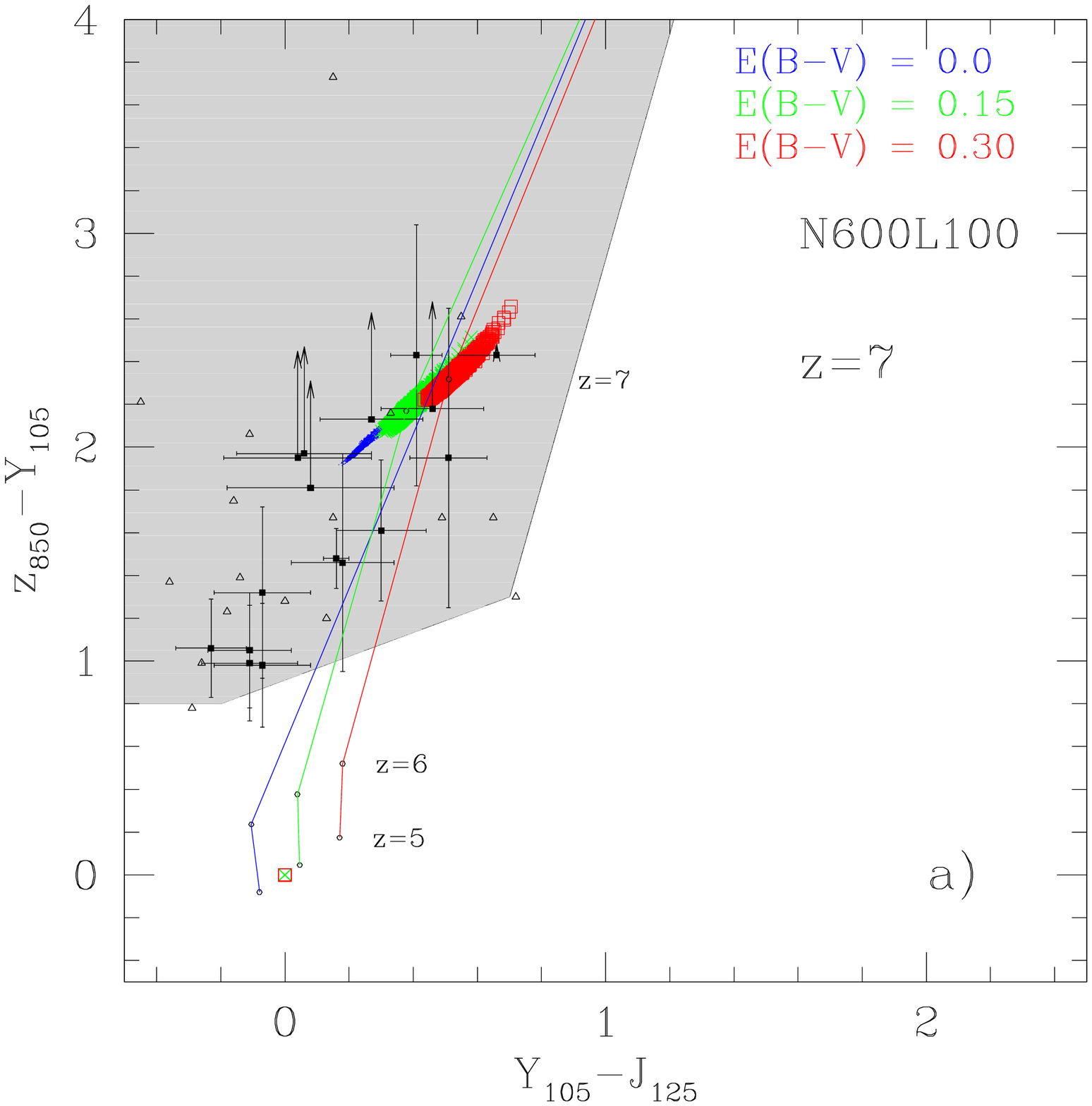}
\includegraphics[width=0.65\columnwidth,angle=0] {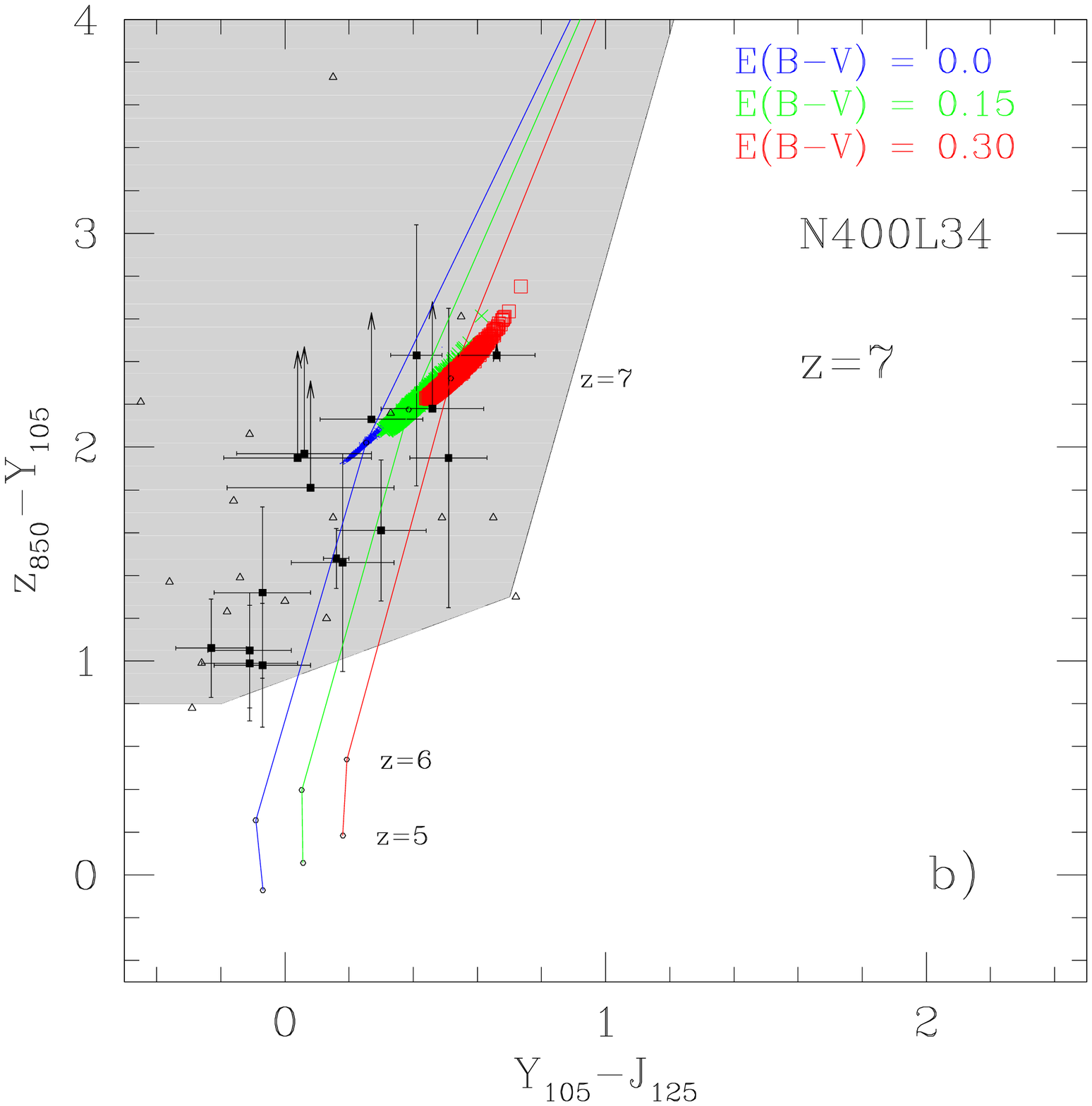}
\includegraphics[width=0.65\columnwidth,angle=0] {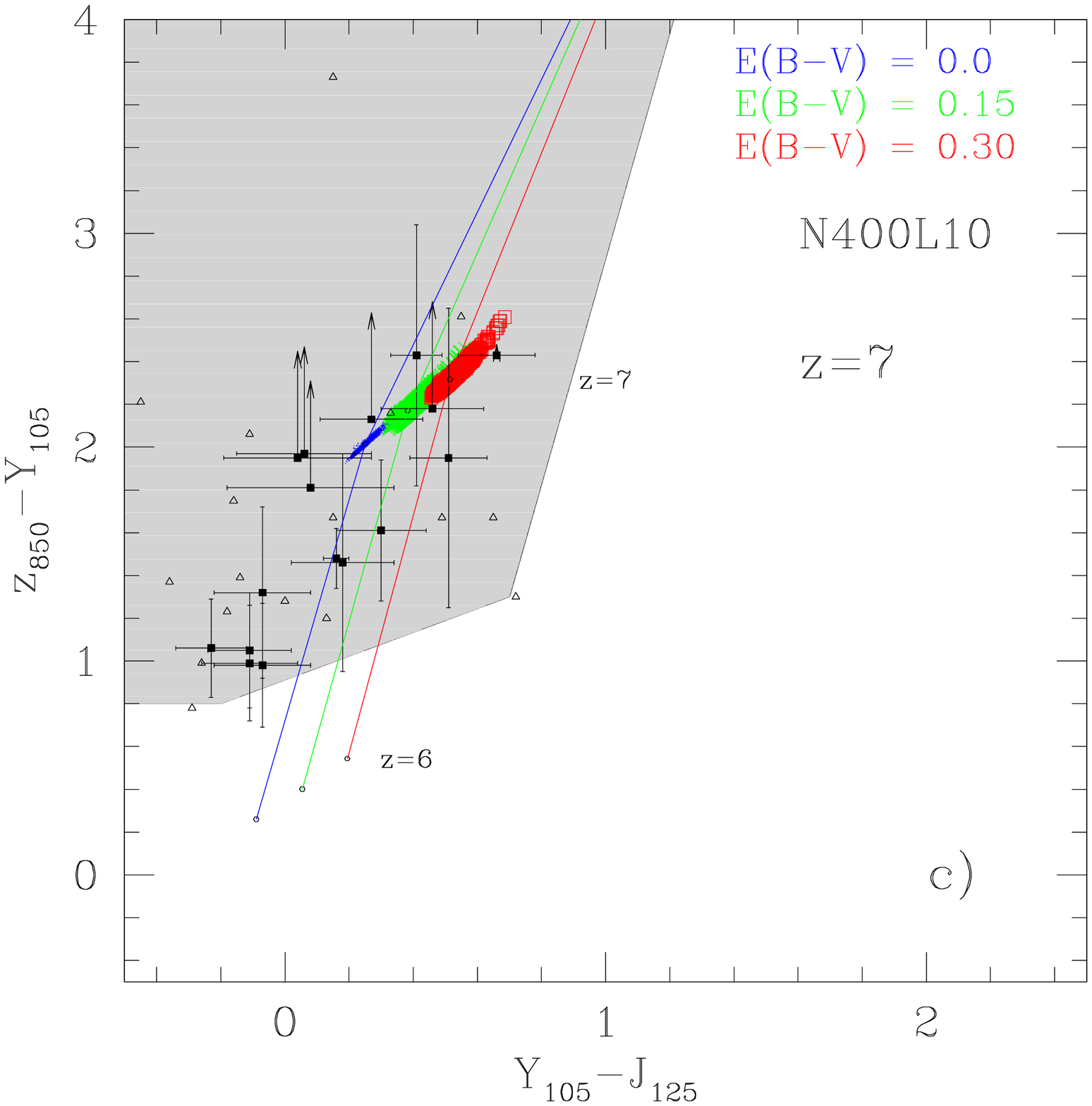}
\caption{
Same as Figure~\ref{fig:ccz6} but for $z\sim 7$. 
The grey shaded area, defined by \citet{Oesch.etal:09}, indicates the expected position of $z\sim 7$ objects.  
For each extinction, the color track of a representative galaxy is shown with a solid line from $z=5$ through $z=8$ (out of plot range).  Black filled squares \citep{Oesch.etal:09} and open triangles \citep{Yan&Windhorst.etal:09} indicate observed $z\sim 7$ candidates. 
}
\label{fig:ccz7}
\end{figure*}

\begin{figure*}
\includegraphics[width=0.65\columnwidth,angle=0] {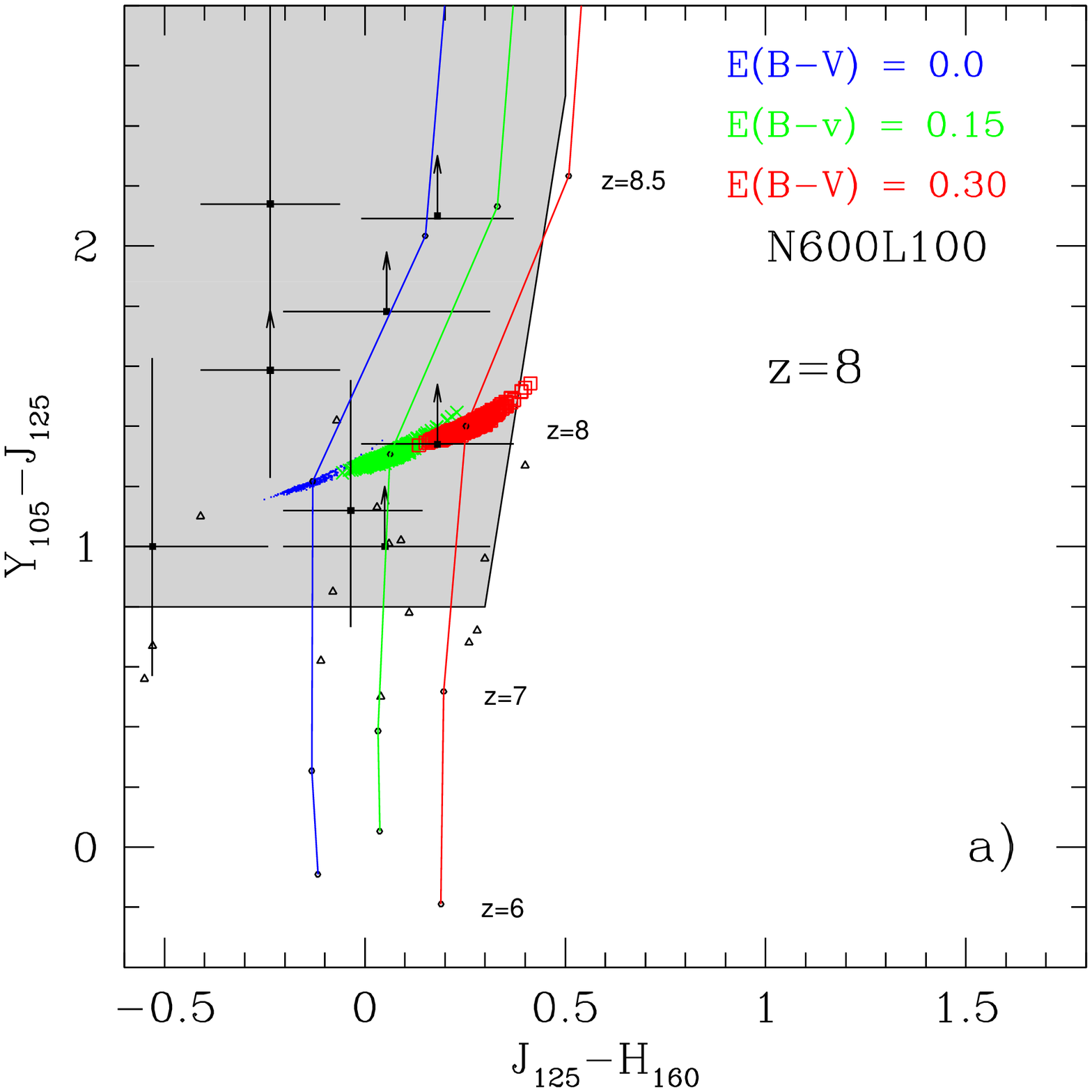}
\includegraphics[width=0.65\columnwidth,angle=0] {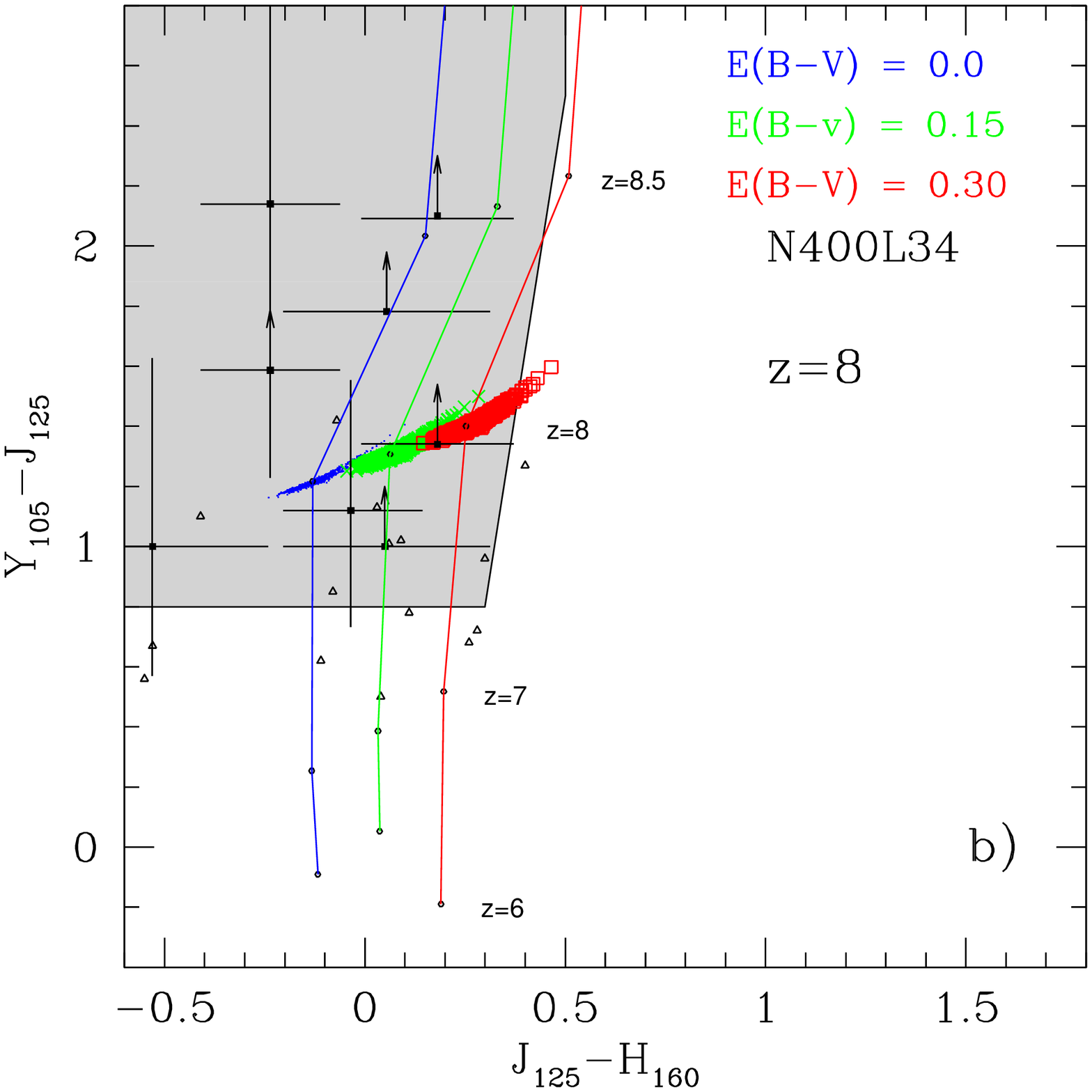}
\includegraphics[width=0.65\columnwidth,angle=0] {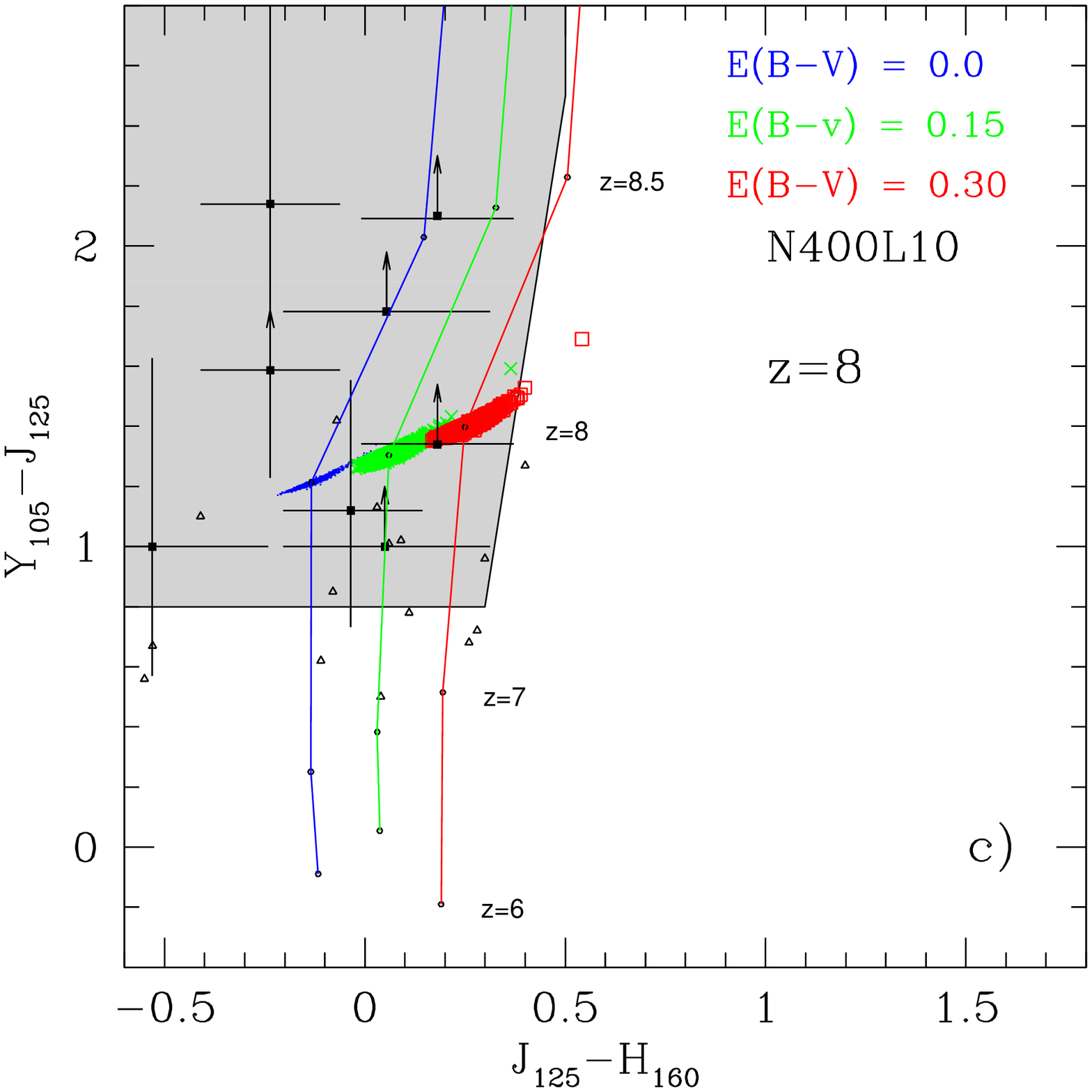}
\caption{
Same as Figure~\ref{fig:ccz6} but for $z\sim 8$.  The grey shaded area, defined by \citep{Bouwens.etal:10a}, indicates the expected position of $z\sim 8$ objects.  For each extinction, the color track of a representative galaxy is shown with a solid line from $z=6$ through $z=9$ (out of plot range).  Black filled squares \citep{Bouwens.etal:10a} and open triangles \citep{Yan&Windhorst.etal:09}  indicate observed $z\sim 8$ candidates. 
}
\label{fig:ccz8}
\end{figure*}

In Figures~\ref{fig:ccz6}, \ref{fig:ccz7} \& \ref{fig:ccz8}, we present color-color diagrams at $z=6$, 7 \& 8, respectively.   For each redshift  we make direct comparisons to the observed data.  In each figure, the results from three simulations (N400L10, N400L34, N600L100) are shown separately, and the simulated galaxies are represented by blue, green, and red points for $E(B-V)=0.0$, 0.15, \& 0.30, respectively.  Galaxy evolution tracks are shown by the corresponding solid blue, red and green lines, which were produced by redshifting the average spectrum of simulated galaxies.  By doing this we are able to show how the Lyman-break of the galaxy spectrum is passing through the HST filter set (Figure~\ref{fig:fnu}a), transitioning into and out of the drop-out candidate selection area which has been defined by observers.  For example in Figure~\ref{fig:ccz6}, $i_{775}-z_{850}$ for $z=6$ galaxies are substantially redder than at for $z=5$,  creating a threshold above which candidates are selected. 
%%Fig. 2
%%Fig. 3
%%Fig. 4
%%Fig. 5
\begin{figure*}
\begin{center}
\includegraphics[width=0.65\columnwidth,angle=0] {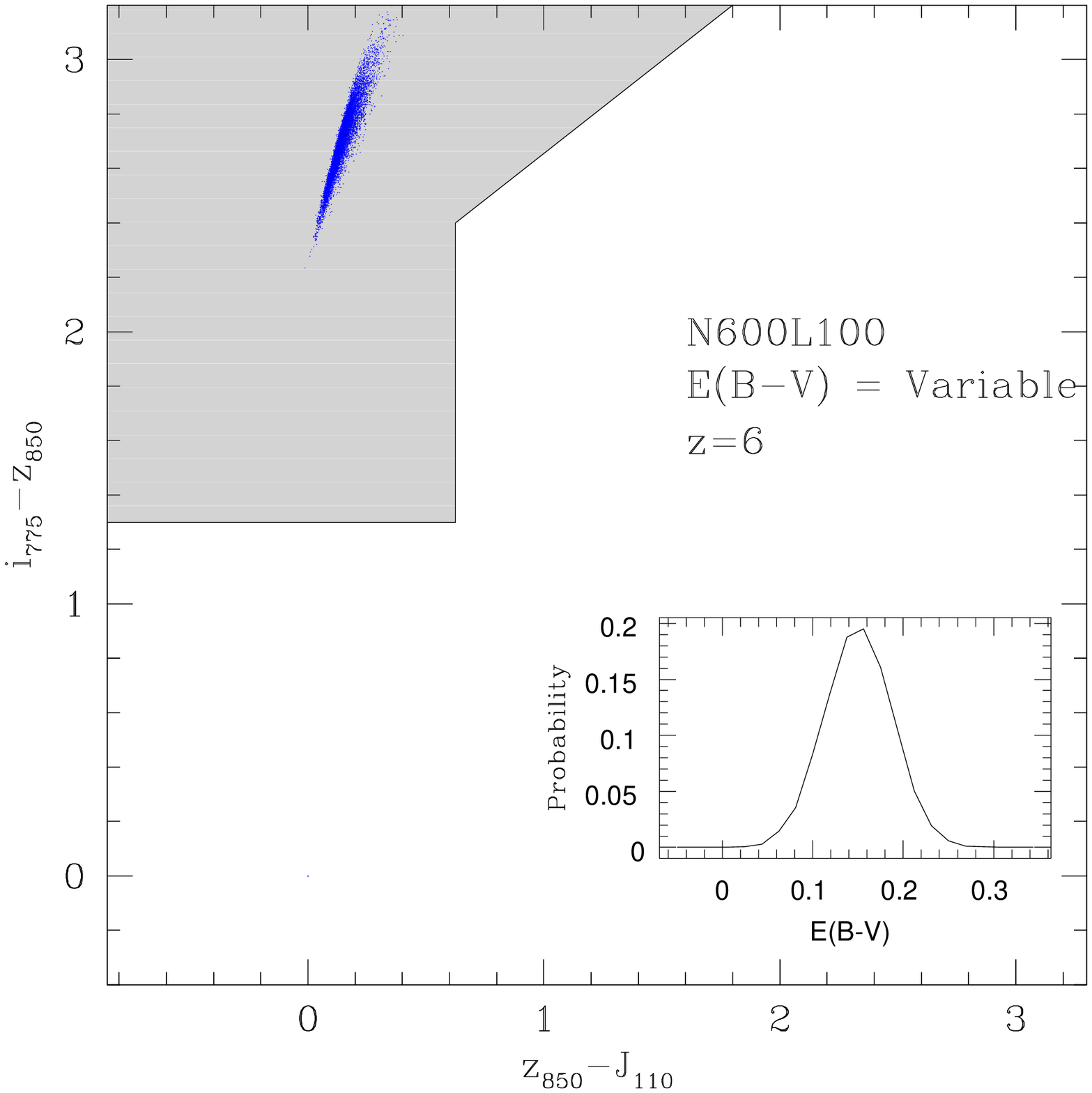}
\includegraphics[width=0.65\columnwidth,angle=0] {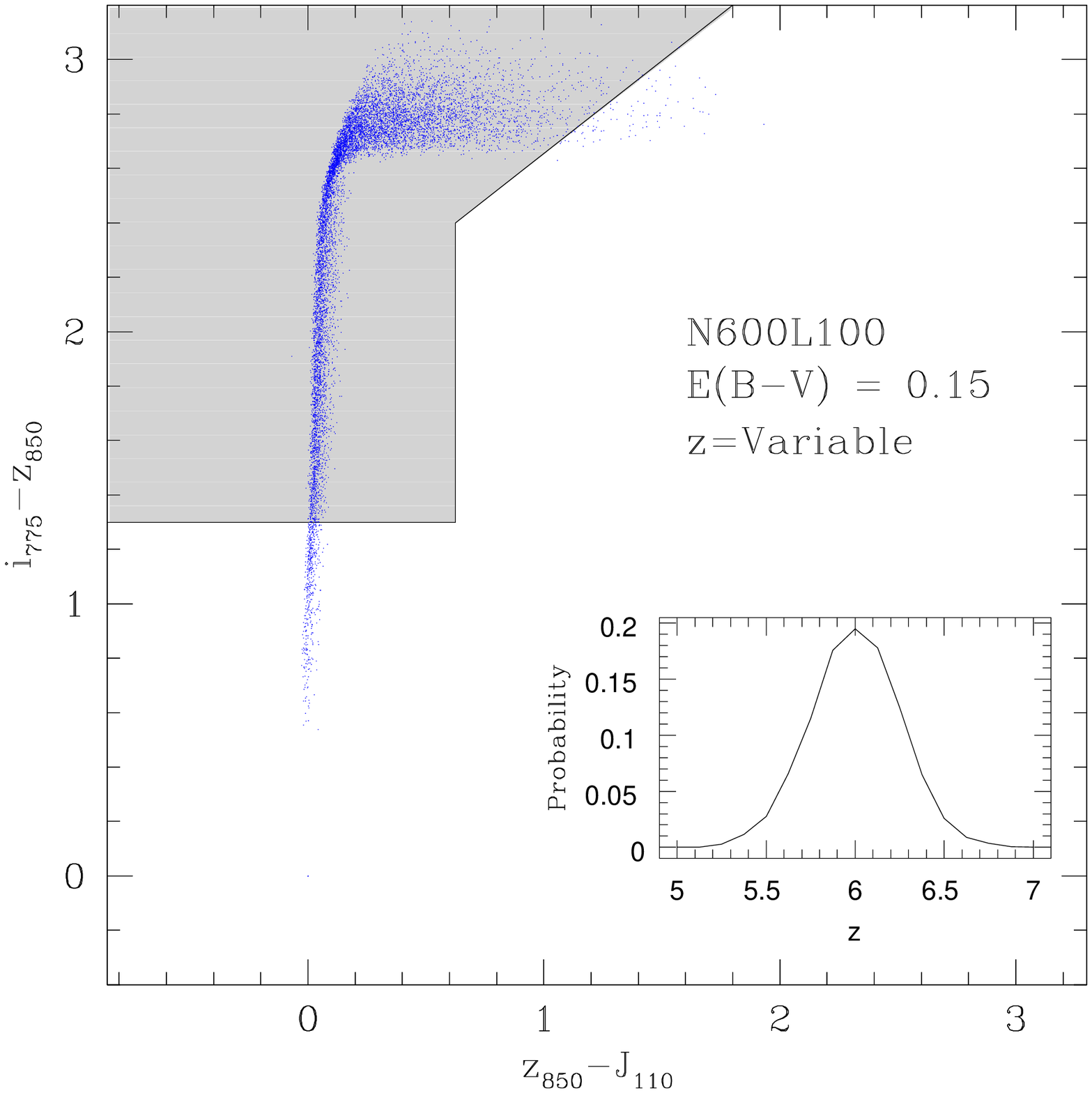}
\includegraphics[width=0.65\columnwidth,angle=0] {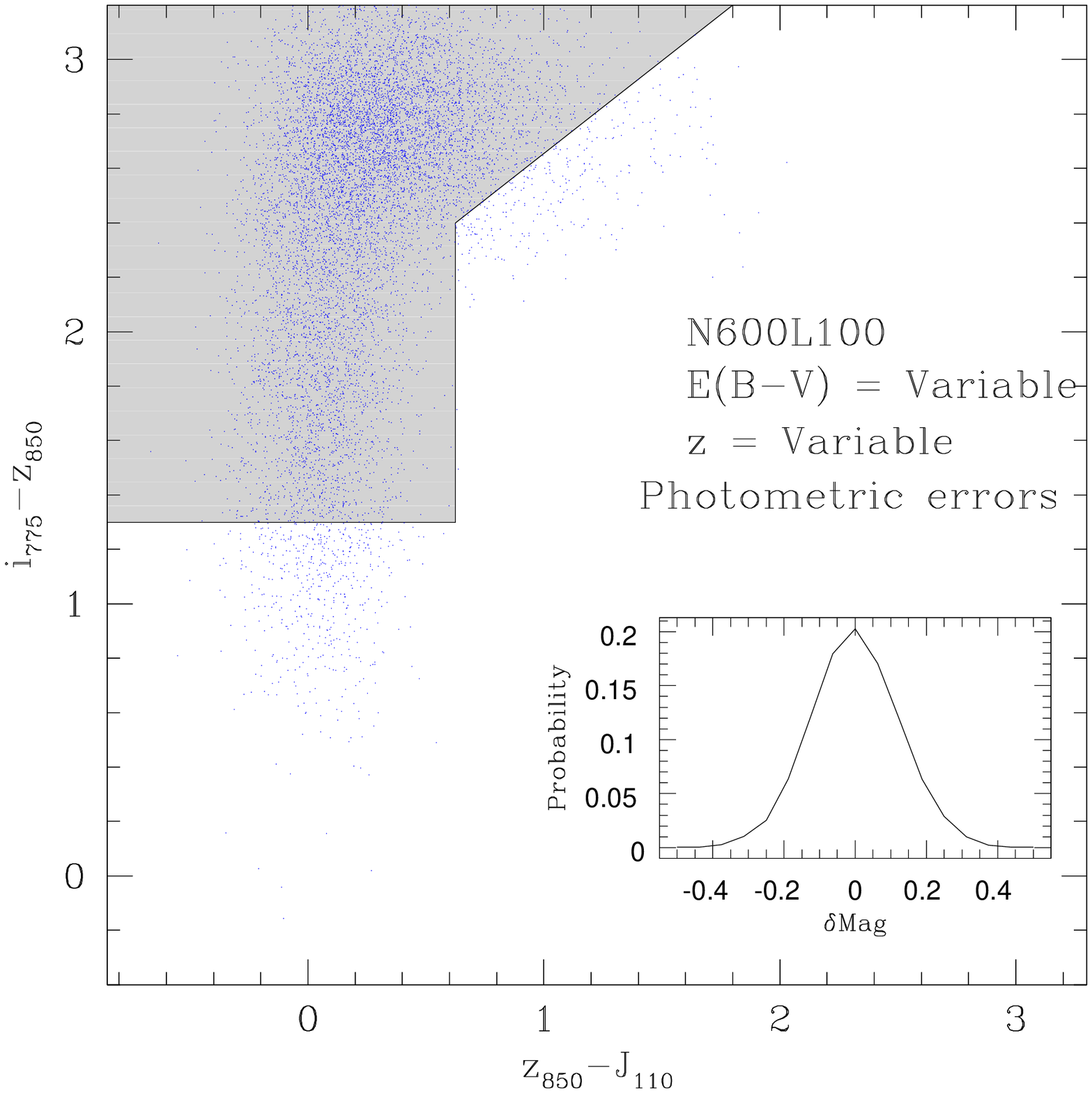}
\caption{This figure explores the origin of scatter in the color-color diagrams. 
{\it Panel (a)}  shows the effect of varying dust extinction.  Inset plot shows a gaussian distribution of E(B-V) around a center value of $0.15$ with a $\sigma=0.038$.  
{\it Panel (b)} shows the effect of varying redshift.  Inset plot shows a gaussian redshift distribution around $z=6$ with a $\sigma=0.25$.
{\it Panel (c)} shows the effect of varying dust extinction, redshift and adding a $\sigma=0.125$ gaussian distribution of photometric errors to the AB magnitudes in each filter.   Inset plot shows distribution of AB magnitudes error.
}
\label{fig:cc_comp}
\end{center}
\end{figure*}

In Figure~\ref{fig:ccz6}, observed candidate data points are represented by solid black squares \citep{Bouwens.etal:06} and open black triangles \citep{Yan&Windhorst:04}.  The grey shaded area indicates the color-color selection criteria defined by \citet{Bouwens.etal:06}.  Due to the abundance of observed points at this epoch, error bars were omitted for clarity. 

Figure~\ref{fig:ccz7} shows color-color plots for $z\sim 7$  with observed data points from \citet[][black filled squares]{Oesch.etal:09}  and \citet[][open triangles]{Yan&Windhorst.etal:09}.  The grey shaded area for color-color selection is taken from \citet{Oesch.etal:09}.  Error bars were included only for the \citet{Oesch.etal:09} data for clarity.  Again we see good agreement with observed data.
 
Figure~\ref{fig:ccz8} shows the color-color diagram for $z\sim 8$ with observed data points from \citet[][black filled squares]{Bouwens.etal:10a} and \citet[][open triangles]{Yan&Windhorst.etal:09}.   Expected position for $z\sim 8$ objects is defined by the grey shaded area \citep{Bouwens.etal:10a}.  Error bars were included only for the \citet{Bouwens.etal:10a} data for clarity.  Here too simulated galaxies show good agreement with observations.

In all of Figures~\ref{fig:ccz6}, \ref{fig:ccz7}, \& \ref{fig:ccz8}, massive galaxies are on the lower left side of the data distribution for the simulated galaxies, because higher mass galaxies have higher SFR in our simulations and hence bluer. 

In Figure~\ref{fig:cc_comp} we examine the tight grouping of galaxies in our color-color plots.  Figure~\ref{fig:cc_comp}a shows the effect of adding a $\pm0.15$ scatter with a $\sigma=0.038$ gaussian distribution centered around $E(B-V)=0.15$.  This addition does little to increase the scatter of the galaxies in color-color space.  In Figure~\ref{fig:cc_comp}b we demonstrate the effect of uncertainties in photometric redshift estimation by adding a $\pm0.50$ scatter with a $\sigma=0.25$ gaussian distribution centered on $z=6$.  This addition produces a vertical feature which runs throughout the selection box with some horizontal scatter at its top.  
Finally in Figure~\ref{fig:cc_comp}c we add a $\pm0.20$ scatter with a $\sigma=0.125$ gaussian distribution to the calculated UV magnitudes as well as including the previous scatter in $E(B-V)$ and redshift.  In this case, we reproduce a similar degree of scatter to the observed data. 
Our results indicate that the current observational color-color selection region can select galaxies with a wide range of extinction values of $E(B-V)=0.0 - 0.30$, and that the scatter in the data points are primarily dominated by the redshift scatter and photometric errors rather than the extinction variance.
The tight grouping of galaxies seen in each of the color-color plots is because the simulated galaxies are extracted at exactly $z=6,7,8$ and have no errors included in magnitude calculations.

%%%%%%%%%%%%%%%%%%%%%%%%%%%%%%%%%%%%%%%%%%%%%%%%%%

\section{Luminosity Functions}
\label{sec:LFs}
%%Fig. 6
\begin{figure*}
\centerline{\includegraphics[scale=0.6] {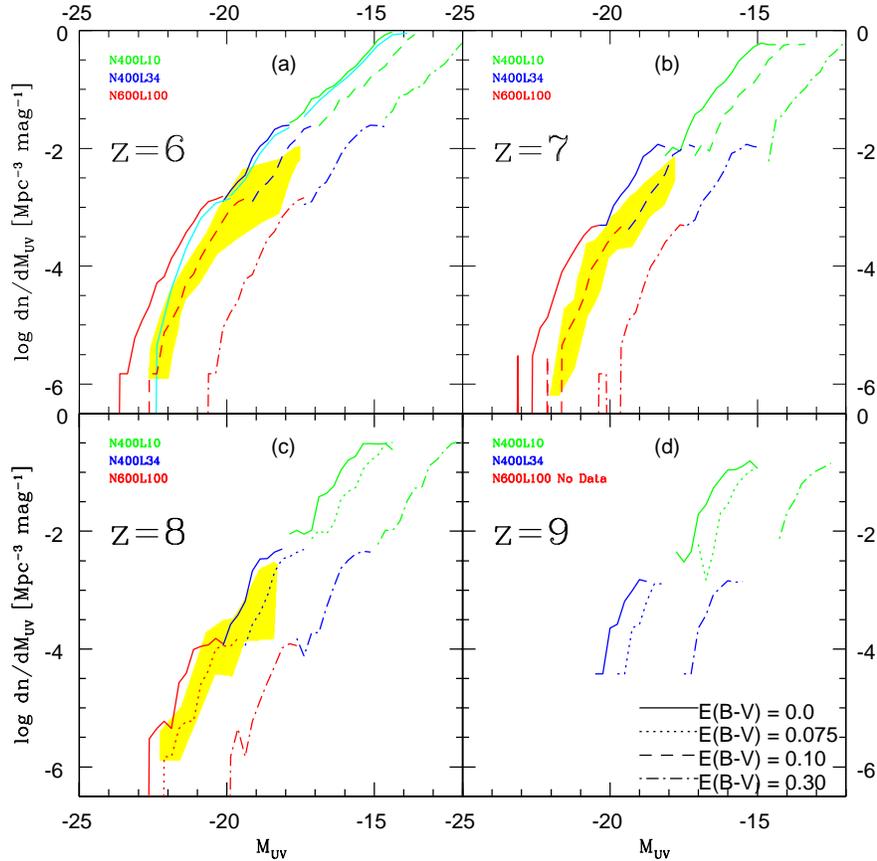}}
\caption{
Composite LFs at $z=6-9$ using the N400L10 (green), N400L34 (blue), N400L100 (red) runs, 
as well as multiple extinction values of $E(B-V)=0.0$ (solid lines), 0.075 (dotted lines), 0.10 (dashed lines), and 0.30 (dot-dashed).  The solid cyan line represents a variable model for dust extinction \citep{Finlator.etal:06}.   For $z=6$. the yellow shade represents a range of observations by \citep{Yan&Windhorst:04,Bouwens.etal:04,Dickinson:04,Malhotra:05,Bunker.etal:04,Beckwith.etal:06}.  
For $z=7$, the yellow shaded area represents a range of observations by \citep{Bouwens.etal:08,McLure.etal:10,Oesch.etal:10b,Oesch.etal:09,Mannucci.etal:07,Richard.etal:06} assembled in \citep{Bouwens.etal:10b}. For $z=8$, observed range represented in yellow is taken from \citep{Bouwens.etal:10b} and includes data from \citep{McLure.etal:10,Castellano.etal:10,Bouwens.etal:10b} and there are no observed data points for $z=9$.}
\label{fig:LF}
\end{figure*}

%%Fig. 7
\begin{figure*}
\begin{center}
\includegraphics[scale=0.6] {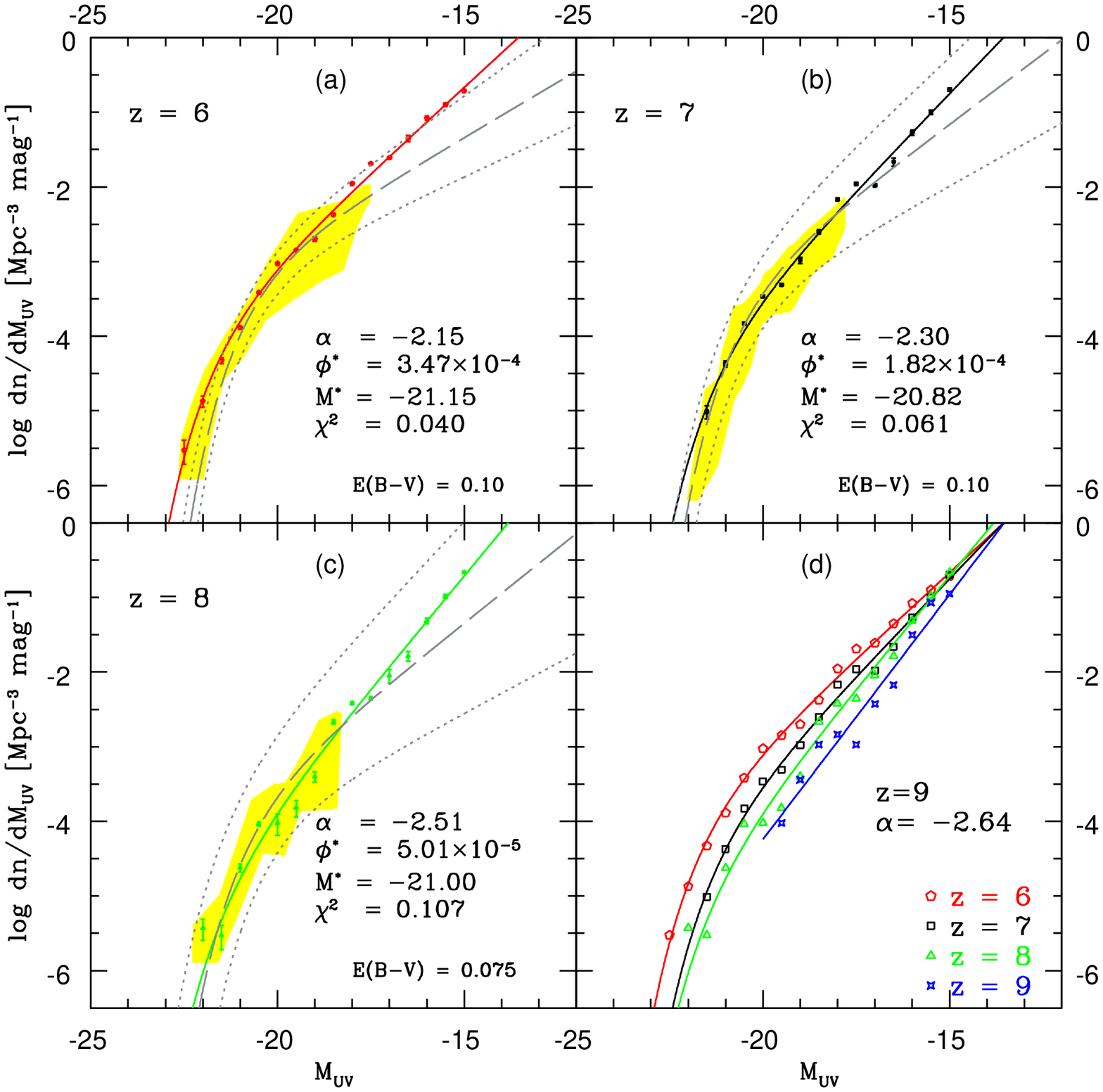}
\caption{Best-fit Schechter LFs at $z=6-9$ for the simulated galaxies.  
The simulation LFs (same as Figure~\ref{fig:LF}) are shown by data points 
together with Poisson error bars. 
The yellow shaded areas for observed data are the same as in Figure~\ref{fig:LF}.  
Schechter best-fits to the observed data \citep[][dashed lines]{Bouwens.etal:10b} is shown together with the $1$-sigma error range (dotted grey lines).  
}
\label{fig:LFsch}
\end{center}
\end{figure*}

The rest-frame UV LFs of simulated galaxies at $z=6-9$ are shown in Figure~\ref{fig:LF}.  To represent a larger range of magnitudes, we construct composite LFs using three simulations with different box sizes, and connect the results at the points where they overlap.  At the brightest end, the data correspond to the last bin in which we have data in the N600L100 run.  At the faintest end, the  minimum $\Muv$ corresponds to the minimum $M_s$ allowed by the mass resolution in the N400L10 run (see Section~\ref{sec:MFs}).  This also corresponds to the point at which the LF begins to turn over, indicating the resolution limit.  For example, at $z=6$, the magnitude coverage of each run are approximately from $\Muv \approx -23$ to $-20$\,mag (N600L100), $-20$ to $-17.5$\,mag (N400L34), and $-17.5$ to $-15$\,mag (N400L10).  These values vary slightly with redshift.   In all panels, three sets of results for $E(B-V)=0.0, 0.10, 0.30$ at $z=6$ \& 7, and $E(B-V)=0.0, 0.075, 0.30$ at $z=8$ \& 9 are shown.  See Section~\ref{sec:method} regarding these extinction choices. 

To explore the impact of a varying extinction, we utilize a model presented in \citet{Finlator.etal:06}: $E(B-V) = 9.0\times Z^{0.9} + \delta E$, where $Z$ is the galaxy stellar metallicity and $\delta E$ is a Gaussian scatter.
This relationship is based on the mean trend between metallicity and $E(B-V)$ found in SDSS at $z\sim 0$. 
In Figure~\ref{fig:LF}(a) the result of this relationship applied to our simulated galaxies is shown by the cyan solid line.  This relationship gives higher extinction for more massive galaxies, therefore our data is consistent with observations only at the very brightest end of the LF with this extinction model. 

The disagreement at fainter magnitudes between the cyan line and the yellow observed region has several potential explanations.  The Finlator's model for $E(B-V)$ is based on the observed relationship at $z\sim 0$, and thus it may not apply at high redshifts.   This model also utilizes metallicity taken from star particles, which is set to the same value as the parent gas particle at the time of formation.  Therefore the metallicity of star particles in our simulation does not evolve with redshift.  We have tested this model utilizing the gas particle metallicity, and found that this change results in lower extinction values, however these values may be artificially lower since this mean metallicity is obtained by averaging over the gas particles including those at the outskirts of galaxies. 
Finally there is the possibility that if high redshift galaxies are mostly dust free as suggested by recent observations \citep{Bouwens.etal:09,Bouwens.etal:11a}, then our simulations are systematically overproducing galaxies by $\sim  0.5$ dex.  This could be true if molecular gas cooling plays an important role in early star formation (see Section~\ref{sec:h2} for further discussion).   However the consistent overproduction of galaxies is not a feature found in our GSMF, as we will discuss further in Section~\ref{sec:MFs} and Figure~\ref{fig:muv_ms}. 

At $z=6$ (top left panel of Figure~\ref{fig:LF}), the yellow shade represents a range of  observations by \citet{Yan&Windhorst:04,Bouwens.etal:04,Dickinson:04,Malhotra:05,Bunker.etal:04,Beckwith.etal:06}.  At this redshift we show good agreement with observed data.

At $z=7$, the yellow shaded area represents a range of observations by \citet{Richard.etal:06, Mannucci.etal:07, Bouwens.etal:08, Oesch.etal:09, McLure.etal:10, Oesch.etal:10b}.  Here too we see good agreement with observations for the N400L34 and N600L100 run.

At $z=8$, the yellow shaded area represents a range of observations by \citet{McLure.etal:10,Castellano.etal:10,Bouwens.etal:10b}.  Again our simulation shows good agreement with observations.

At $z=9$, we only show the results from the N400L10 and N400L34 runs as there are no objects found at this redshift in the N600L100 run due to low resolution.  Also there has been no published data for robust observed LF estimates at $z=9$.

%%%%%%%%%%%%%%%%%%%%%%%%%%%%%%%%%%%%%%%%%%%%%%%%%%

\subsection{Schechter Function Fit}
\label{sec:SSF}

We perform $\chi^{2}$ fit analysis of the \citet{Schechter:76} function to the composite simulated galaxy LFs.  The Schechter function, $\phi(\Muv)\equiv dn/d\Muv$, as a function of magnitude is given by
%\begin{equation}
\begin{align}
\label{eq:schlum}
\nonumber \phi(\Muv)=\phi_L^{*}(0.4\ln 10) 10^{0.4(\Muv^{*}-\Muv)(1+\alpha_L)} \\ 
\exp(-10^{0.4[\Muv^{*}-\Muv]}),
\end{align}
%\end{equation}
where $\phi^{*}$, $\Muv^{*}$ and $\alpha_L$ are the normalization, characteristic magnitude, and faint-end slope.

We minimize the expression 
\begin{equation}
\label{chi}
\chi^{2} =\displaystyle\sum\limits_{i=0}^{N_{\rm bins}}\frac{[O_{i}-E_{i}]^{2}}{|E_{i}|}
\end{equation} 
by systematically adjusting each of the three Schechter parameters, where $O_{i}$ is our simulated data set, $E_{i}$ is the calculated value of the above Schechter function, and $N_{\rm bins}= 16, 14, 15,$ \& 10 for $z=6,7,8,$ \& 9, respectively.  For $z=9$ a simple power-law was used, because the N600L100 run did not contain any galaxies at this redshift due to limited resolution. The $1$-sigma error of the parameters are determined by fixing two of the best-fit parameters  and varying the third from $\chi^{2}_{\rm min}$ to $\chi^{2}_{\rm min}+1$ \citep{Andrae:10}.

%%Table 2
\begin{table}
\begin{center}
\begin{tabular}{ccccc}
\hline
 $z$ & $\alpha_L$ & $\log( \phi^*)$ & $\Muv^*$ & $\chi^{2}$  \\ 
\hline
$6$ & $-2.15^{+.24}_{-.15}$ & $-3.46^{+.29}_{-.37}$ & $-21.15^{+.53}_{-.53}$ & $0.04$\vspace{0.2cm} \\
$7$ & $-2.30^{+.28}_{-.18}$ & $-3.74^{+.32}_{-.41}$ & $-20.82^{+.61}_{-.56}$ & $0.06$\vspace{0.2cm} \\
$8$ & $-2.51^{+.27}_{-.17}$ & $-4.30^{+.32}_{-.42}$ & $-21.00^{+.59}_{-.50}$ & $0.11$\vspace{0.2cm} \\
$9$ & $-2.64^{+.12}_{-.13}$ & --- & --- & ---\\
\hline
\end{tabular}
\caption{Best-fit Schechter LF parameters at $z=6-9$ for the simulated galaxies.
Some parameters at $z=9$ are unavailable due to lack of data at the bright end from the N600L100 run. }
\label{tbl:sch}
\end{center}
\end{table}

Figure~\ref{fig:LFsch} shows our $\chi^{2}$ best-fits to the Schechter function along with the observational fits (long-dashed line) by \citet{Bouwens.etal:10b}.  The dotted line represents the $1$-sigma error range of \citet{Bouwens.etal:10b} fit.  The yellow shaded areas are the same as in Figure~\ref{fig:LF}.  For the Schechter fits, we used the simulation data sets with $E(B-V)=0.10$ for $z=6$ \& 7, and  $E(B-V)=0.075$ for $z=8$ \& 9, as they agree with the yellow shade best. 
The best-fit Schechter parameters are given in Table~\ref{tbl:sch}.

%%Fig. 8
\begin{figure*}
\begin{center}
\includegraphics[width=0.7\columnwidth,angle=0] {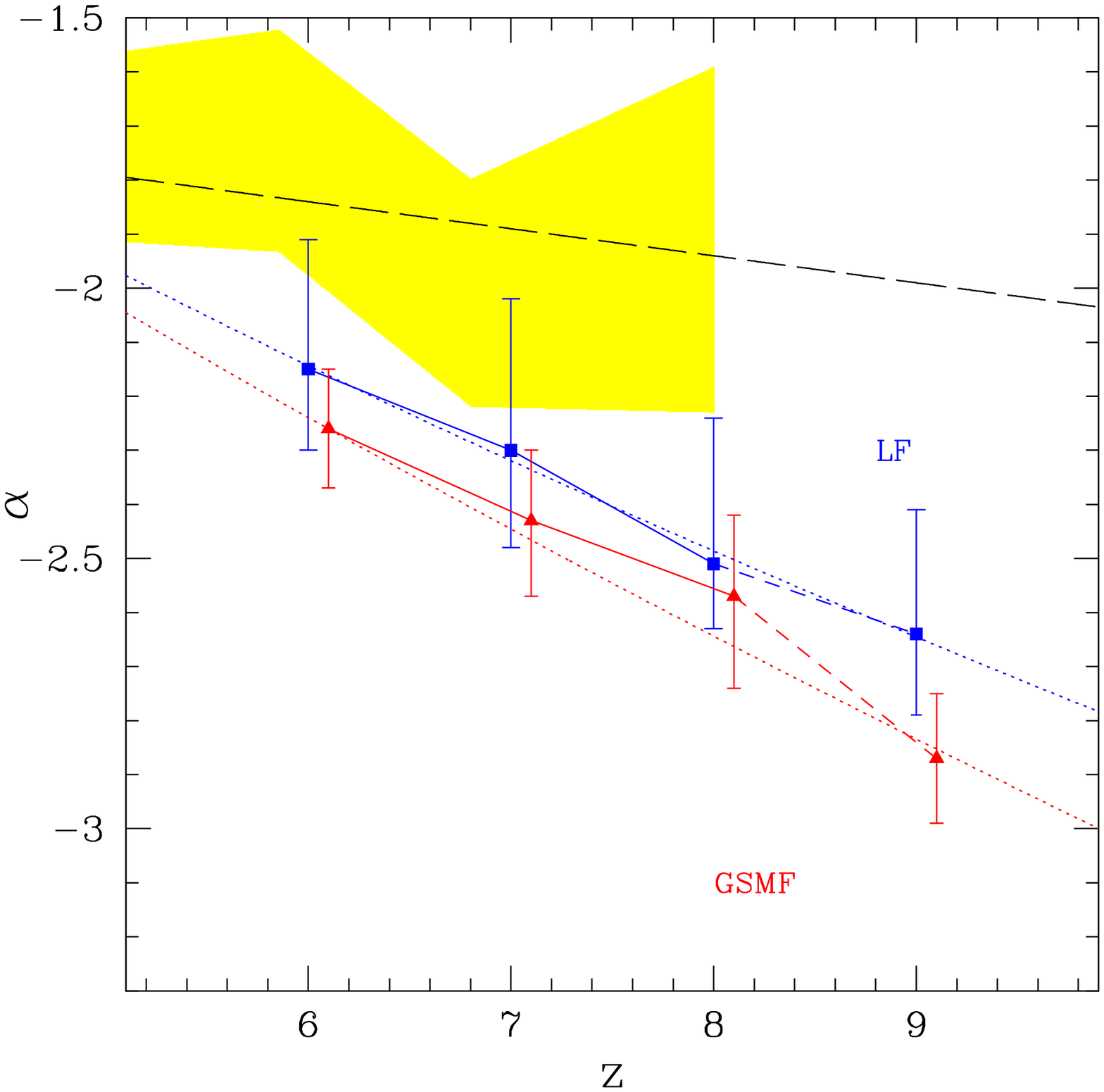}
\includegraphics[width=0.7\columnwidth,angle=0] {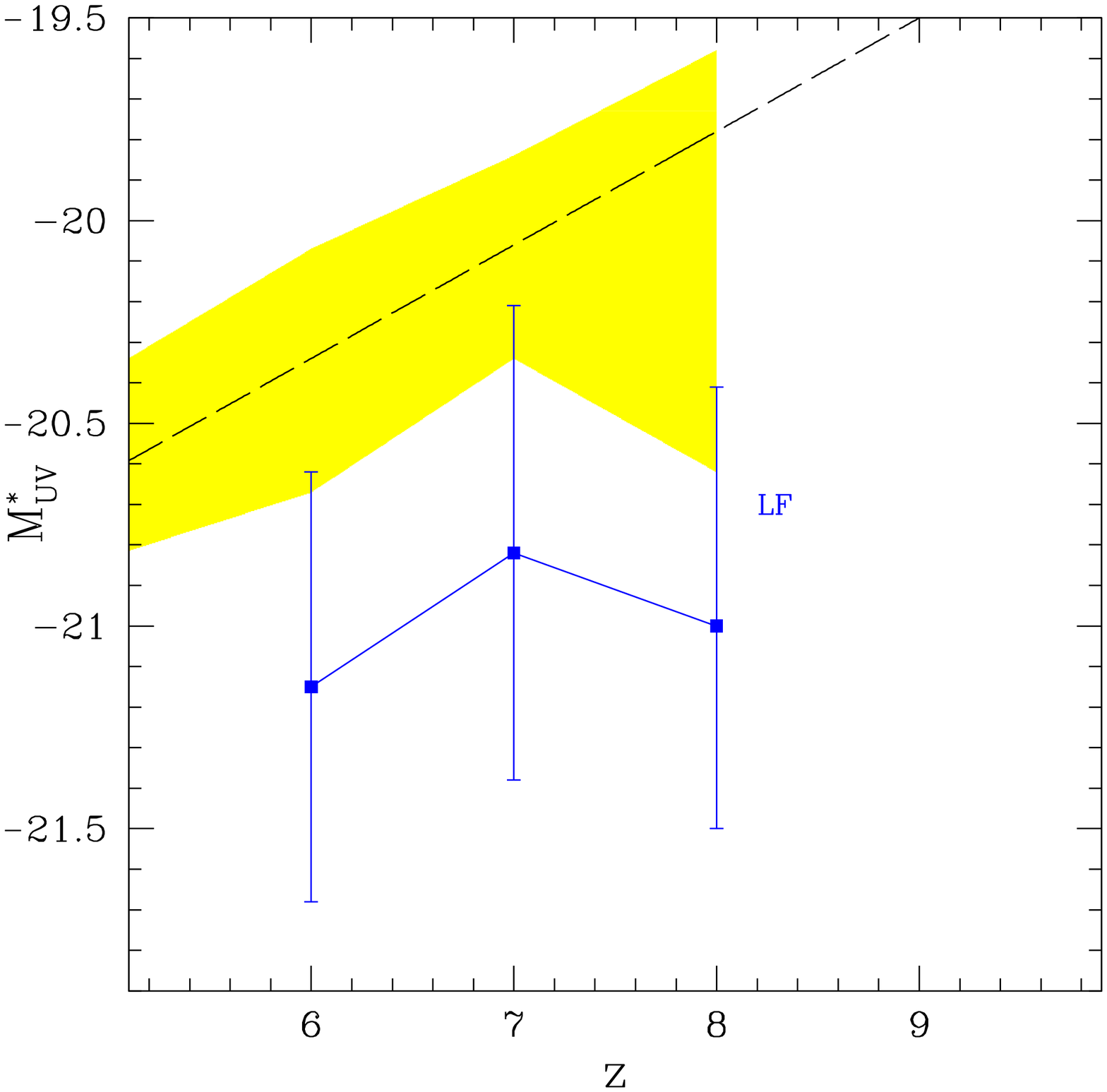}
\includegraphics[width=0.7\columnwidth,angle=0] {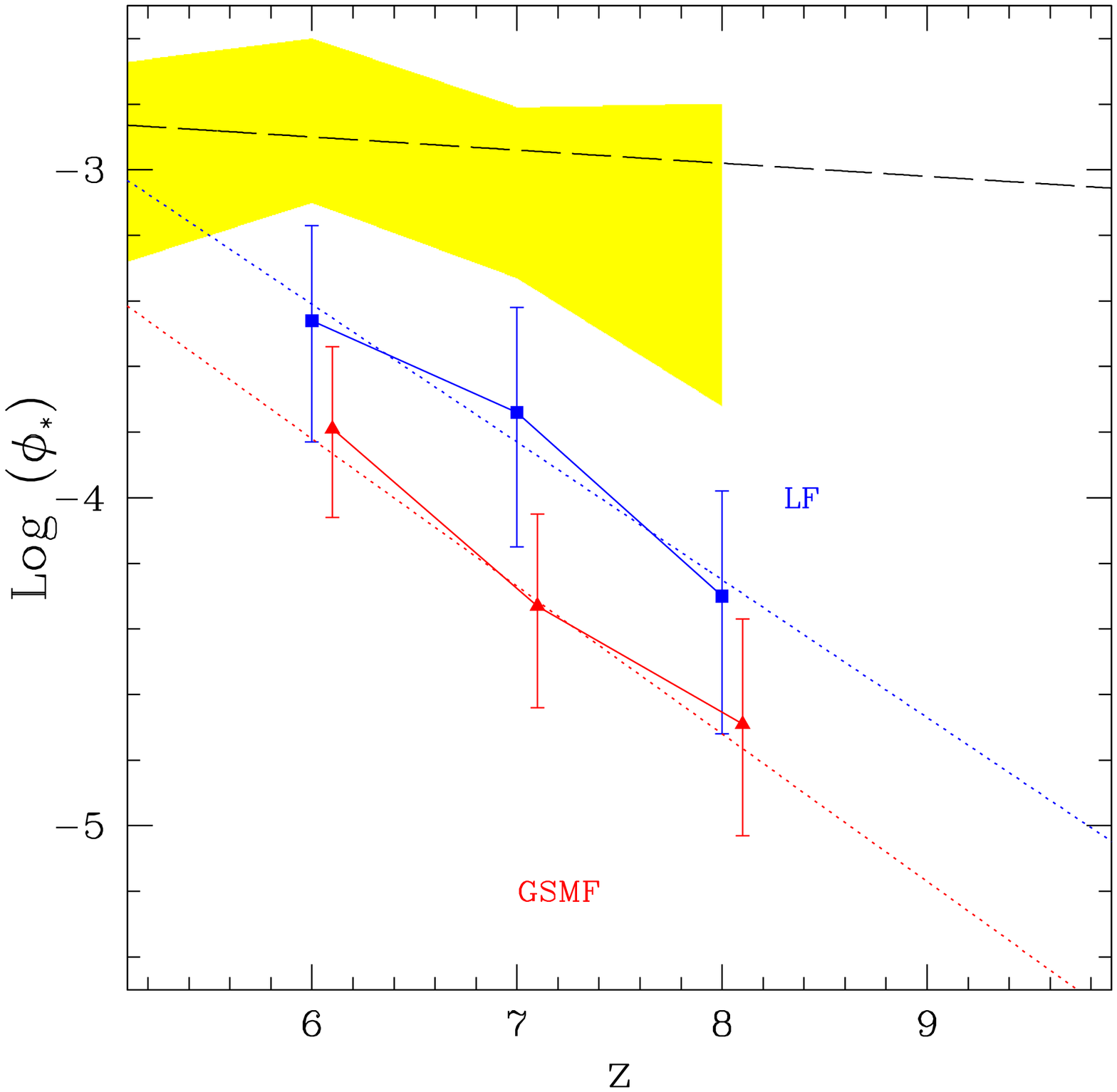}
\includegraphics[width=0.7\columnwidth,angle=0] {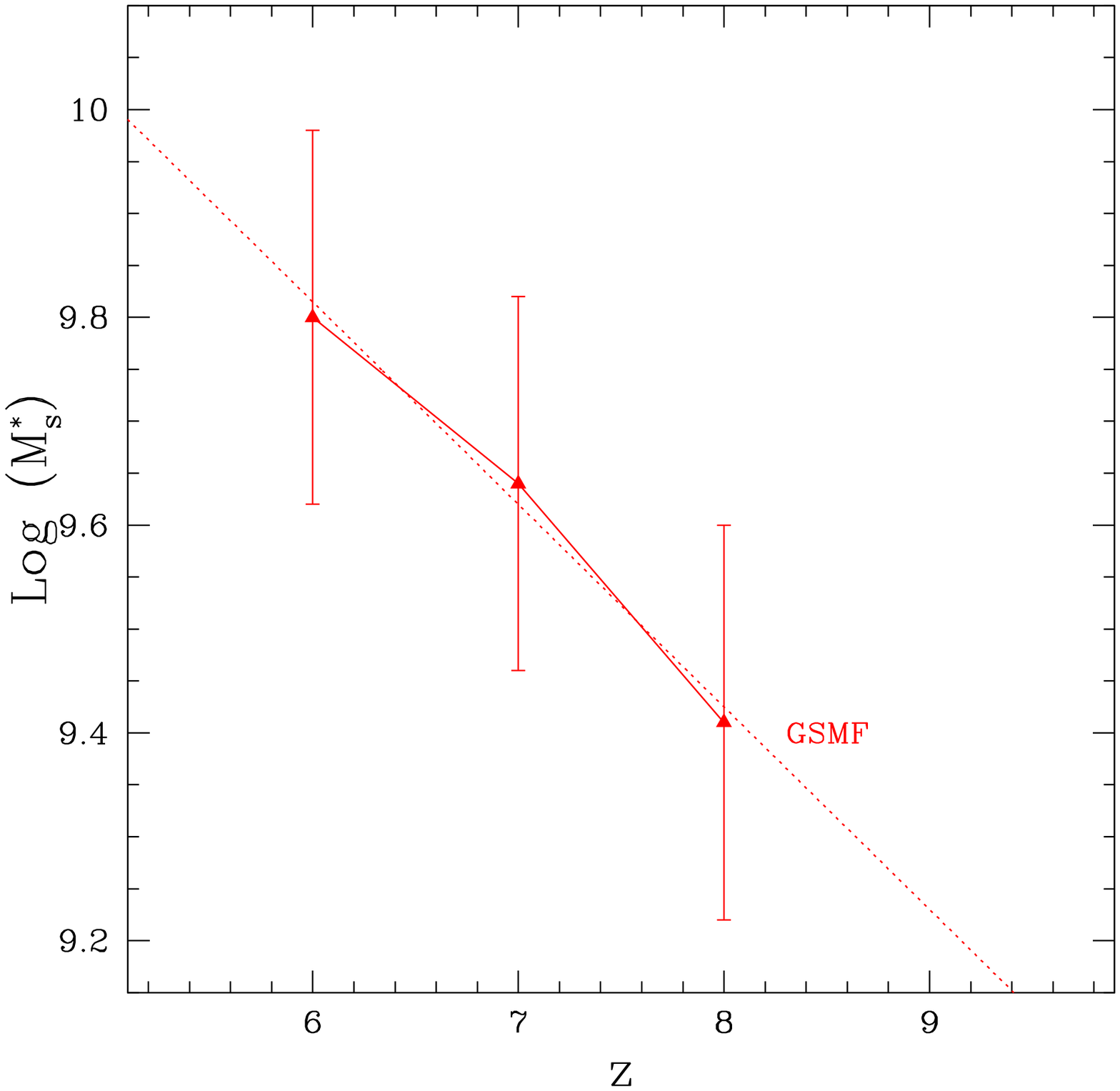}
\caption{
Evolution of Schechter parameters for both LF (blue) and GSMF (red) as functions of redshift.  The dashed line connecting $z=8$ data points to $z=9$ indicates that the $z=9$ data set is incomplete at the bright end. $1$-sigma error bars are shown for all data points, see \ref{sec:SSF} for detail of error analysis.   Red data points are offset for clarity in $\alpha$ and $\phi^*$ plots.  Dotted red and blue lines indicate $(1+z)^\gamma$ fits to the data in each plot with the exception of $\Muv^*$ which shows only a week trend with redshift.   Yellow shaded area represents the observed fit parameter range found in \citet{Bouwens.etal:11a} with the black dashed line indicating the linear fit to that data.
}
\label{fig:fit}
\end{center}
\end{figure*}

Both our Schechter fit and the simulation LFs agree well with the yellow observed range at the bright end.  However our best-fit Schechter LFs are noticeably steeper than the observed one at the faint end.  This is caused by numerous faint galaxies with $M_{\rm UV} > -18$~mag in our simulation, which lie beyond the current detection limit.  Our results suggest that future surveys with increased sensitivity (such as the JWST) will detect a large number of faint galaxies. 
Although at different redshifts, this trend of finding more galaxies below the previous magnitude limit seems to be occurring in recent observations. For example, \citet{Drory.etal:10} found an upturn toward a steep faint-end slope of $\alpha_L \sim −1.7$ at lower mass at $z\le 1$, when they probed down to fainter limits than before in the COSMOS field. 
Also, \citet{Reddy.etal:10} found a steeper slope of $\alpha_L=-1.74$ at $z\sim 2-3$ than before using a larger sample of $>2000$ spectroscopic redshifts and $\sim$\,31000 LBGs. 

We summarise the evolution of Schechter parameters of simulated galaxies in Figure~\ref{fig:fit}. The yellow shades show the range of observed data points taken from \citet{Bouwens.etal:10a}. 
In the top left panel, our simulation shows a clear evolution of $\alpha_L$ from $\alpha_L=-2.15$ at $z=6$ to $\alpha_L=-2.51$ at $z=8$, and our slopes are in general steeper than the current observations by $\Delta \alpha_L \simeq 0.3$.  At $z=9$, we used a power-law fit to the incomplete data set (missing bright-end in the N600L100 run), therefore the $z=8$ \& 9 data points are connected by a dashed line.  

%%Fig. 9
\begin{figure*}
\begin{center}
\includegraphics[scale=0.6] {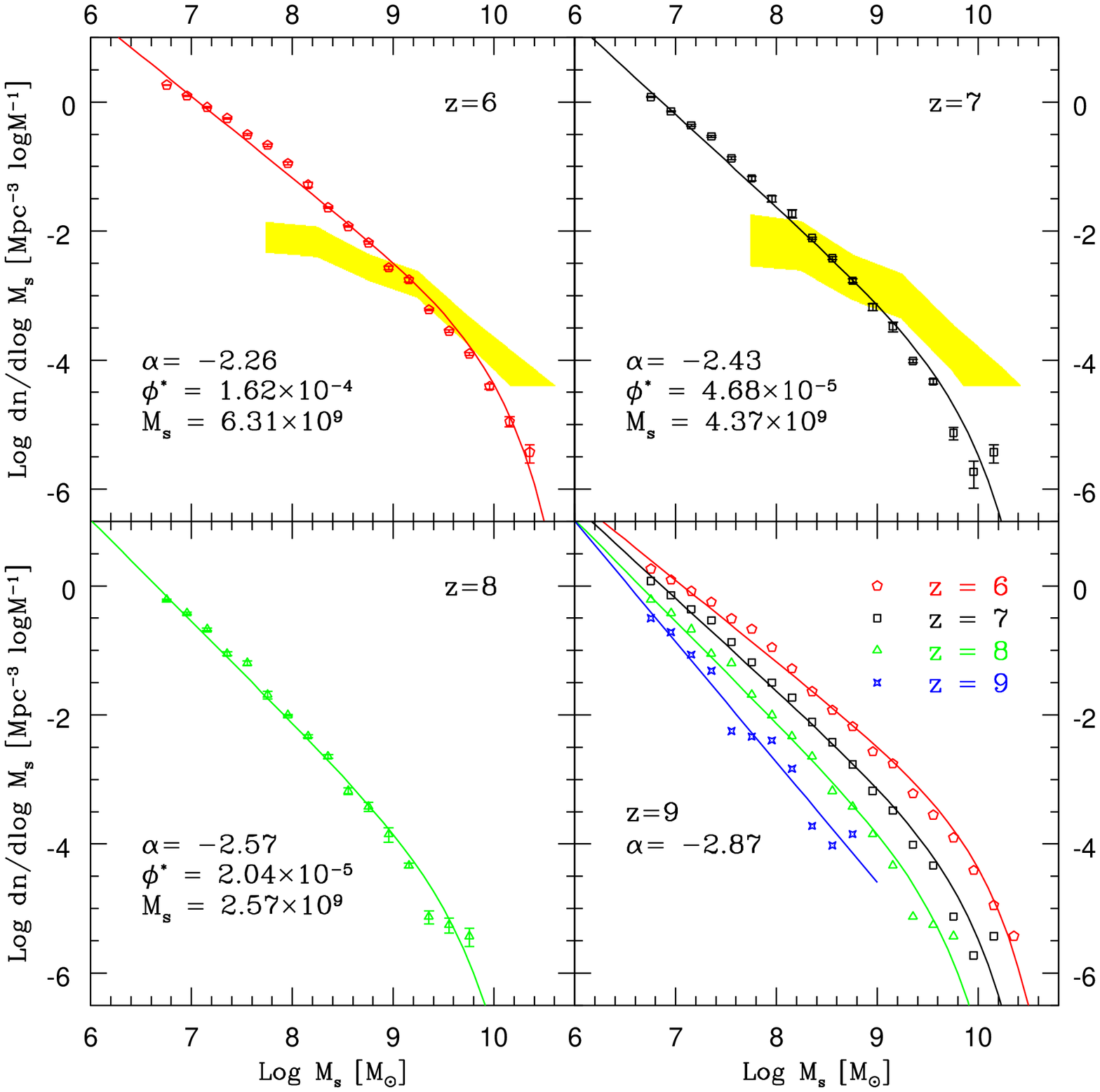}
\caption{
Composite GSMFs at $z=6-9$ with best-fit Schechter LFs.  The best-fit Schechter parameters are summarised in Table~\ref{tbl:schm}.  Yellow shaded regions at $z=6$ and $z=7$ indicate observational derived mass functions taken from \citet{Gonzalez.etal:10} .   
}
\label{fig:massf}
\end{center}
\end{figure*}

Our simulations suggest a steepening trend of 
\begin{eqnarray}
| \alpha_L | &=& 0.68(1+z)^{0.59}, \quad {\rm or}\\
\alpha_L &=&  -0.17(z-6) - 2.16. 
\end{eqnarray}
This steepening of $\alpha_L$ towards higher redshift is in congruence with the result of \citet{Bouwens.etal:11a},  which finds a similar redshift evolution.  
As we will show in the next section, this steepening of $\alpha_L$ is driven by the simultaneous steepening of galaxy stellar mass function. In a $\Lambda$CDM universe, we expect a larger number of lower mass galaxies at higher redshifts, and these low mass galaxies gradually merge to form more massive galaxies. Therefore the steepening of $\alpha$ with increasing redshift is naturally expected in our simulations. 

The characteristic magnitude $\Muv^*$ in our simulation does not show much evolution, except that it brightens by about 0.3 mag from $z=7$ to $z=6$. The simulation $\Muv^*$ is brighter than the observed range by $0.5-1$\,mag. 
The normalization parameter $\phi_L^*$ of simulated galaxies increases significantly by about an order of magnitude from $z=8$ to $z=6$ with a redshift dependence of 
\begin{equation}
\phi_L^* \propto (1+z)^{-0.42}.
\end{equation}
The simulation $\phi_L^*$ is lower than the observed one by about $0.5$ dex.

One of the possible reasons for our lower $\phi_L^*$ and brighter $\Muv^*$ than the observational estimates is that our LF does not have a distinct exponential break at the bright-end, and the simulation LF seems more like a single power-law with a slight curvature at the bright-end. Therefore we are somewhat forcing a Schechter fit, which could push the $\Muv^*$ to brighter side $\phi_L^*$ to lower side. 
Future simulations with AGN feedback may make the bright-end break more pronounced, and the values of above two parameters might approach the observed range. 

%%%%%%%%%%%%%%%%%%%%%%%%%%%%%%%%%%%%%%%%%%%%%%%%%%

\section{Galaxy Stellar Mass Functions}
\label{sec:MFs}

\begin{figure}
\begin{center}
\includegraphics[scale=0.43] {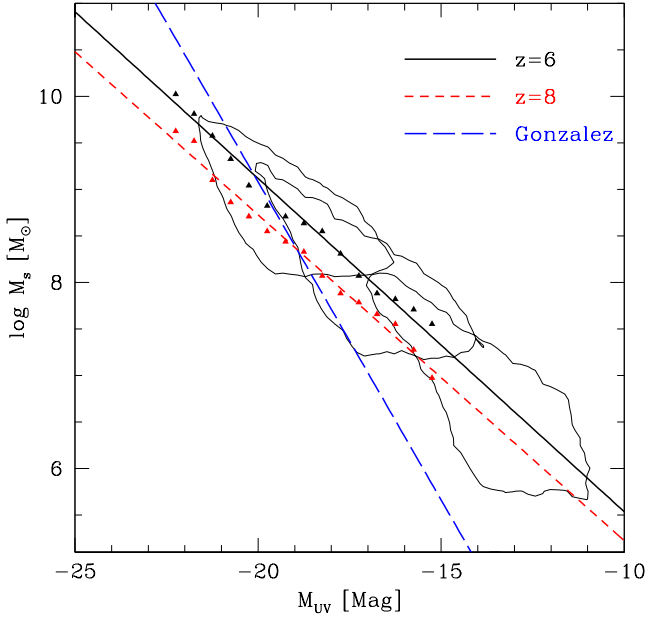}
\caption{
Relationship between rest-frame UV magnitude and galaxy stellar mass in our simulations indicated by the black contours (N600L100, N400L34 and N400L10 from upper left to lower right) for $z=6$.  Median values in each $0.5$\,mag bin are shown by the black and red triangles for $z=6$ and 8, respectively, with a least-square linear fit to the median points (solid black and red short-dashed lines, with a slope of $\approx -0.35$).  Points greater than our $\Muv$ limit of $-15$ are exclude from the fit.  The blue long-dashed line is the observational result from \citet{Gonzalez.etal:10} with a slope of $-0.68$. 
}
\label{fig:muv_ms}
\end{center}
\end{figure}

Galaxy stellar mass functions (GSMFs) in our simulations at $z=6-9$ are shown in Figure~\ref{fig:massf}.  As with the luminosity function, data from three simulations have been combined to show a continuous mass function.  The yellow shaded area shows the observed mass functions for $z=6$ \& 7 \citep{Gonzalez.etal:10}, which was derived directly from the luminosity function obtained by \citet{Bouwens.etal:10a}.
The lower mass limit of $M_s = 10^{6.8} \Msun$ for the N400L10 run was determined by converting the faint-end luminosity cut-off of $\Muv = -15$ to $M_s$ via the relationship between $\Muv$ and $\log(M_s)$ in our simulations (Figure~\ref{fig:muv_ms}).  In order to stay within the mass resolution limits of each run, we take the following mass range from each run:  $M_s= 10^{6.8}-10^8 \Msun$ (N400L10), $10^8-10^9 \Msun$ (N400L34), $>10^9 \Msun$ (N600L100).  
This ensures that, based on the star particle mass for each run (Table~\ref{tbl:Sim}), we will have a sufficient number of particles in each galaxies.  For example, the N400L10 run has a star particle mass of $9.55\times 10^4 \Msun$, which is half of the gas particle mass.  Therefore a galaxy of $10^{6.8} \Msun$ would contain approximately $66$ star particles.  Simulated galaxies would also contain gas and dark matter particles, so the actual particle number of simulated galaxies would be significantly larger.  Further discussion of the simulation resolution limits can be found in Section \ref{sec:Res}.

The Schechter function $\phi(M_s)\equiv \ln(10) M_s \Phi(M_s)\equiv dn/d\log M_s$ as a function of galaxy stellar mass $M_s$ is defined as
\begin{equation}
\label{eq:schmass}
\phi(M_s)= (\ln10)\phi_M^*\left( \frac{M_s}{M_s^*}\right)^{(1+\alpha_M)}\exp \left( -\frac{M_s}{M_s^*} \right), 
\end{equation}
where $\phi_M^*$, $M_s^*$, and $\alpha_M$ are the normalization, characteristic stellar mass, and low-mass (faint-end) slope, respectively.   
We perform the same $\chi^{2}$ analysis for these three Schechter parameters.  The best-fit parameters are given in Table~\ref{tbl:schm}.  Similarly to the LF, we find that the GSMF steepens from $\alpha_M = -2.26$ at $z=6$ to $\alpha_M = -2.57$ at $z=8$, with an evolutionary trend of 
\begin{eqnarray}
|\alpha_M | &=& 0.63(1+z)^{0.65}, \quad {\rm or} \\
 \quad \alpha_M &=& -0.20 (z-6) - 2.26.   
\end{eqnarray}

Figure~\ref{fig:fit} includes the results of GSMF Schechter fits, showing a clear trend that galaxies become more massive and numerous from $z=9$ to $z=6$ as they grow with time.  The characteristic number density $\phi_M^*$ and the associated $M_s^*$ increase with decreasing redshift, evolving as
\begin{equation}
\phi_M^* \propto (1+z)^{-0.45},
\end{equation}
and
\begin{equation}
M_s^* \propto (1+z)^{-0.20}.
\end{equation}
These trends are also summarised in the bottom right panel of Figure~\ref{fig:massf}, clearly showing the evolution of GSMF from $z=9$ to $z=6$.    

\begin{table}
\begin{center}
\begin{tabular}{ccccc}
\hline
 $z$ & $\alpha_M$ & $\log(\phi^*)$  & $\log(M_s^*)$ & $\chi^{2}$  \\ 
\hline
6 & $-2.26^{+.11}_{-.11}$ & $-3.79^{+.25}_{-.27}$ & $9.80^{+.18}_{-.18}$ & $0.25$ \vspace{0.2cm} \\
7 & $-2.43^{+.13}_{-.14}$ & $-4.33^{+.28}_{-.31}$ & $9.64^{+.18}_{-.18}$ & $0.30$\vspace{0.2cm} \\
8 & $-2.57^{+.15}_{-.17}$ & $-4.69^{+.32}_{-.34}$ & $9.41^{+.19}_{-.19}$ & $0.18$\vspace{0.2cm} \\
9 & $-2.87^{+.12}_{-.12}$ & --- & --- & ---\\
\hline
\end{tabular}
\caption{Best-fit Schechter parameters for simulated GSMFs at $z=6-9$.
The evolution of these parameters are also summarised in Figure~\ref{fig:fit}.
Some parameters at $z=9$ are unavailable due to lack of data at the massive end from the N600L100 run. }
\label{tbl:schm}
\end{center}
\end{table}

Interestingly the slope $\alpha_M$ is steeper for the GSMFs than for the LFs by $\Delta \alpha \sim 0.1$.  
This can be explained by examining the relationship in our simulations between $\Muv$ and $\log(M_s)$ (Figure \ref{fig:muv_ms}).  Through a linear fit to the median data point in each $\Muv$ bin, we find 
\begin{equation}
\log(M_s)=\beta\,\Muv + C,
\end{equation}
where $\beta\approx -0.35$ and $C=1.90, 1.75, 1.59,1.58$ at $z=6,7,8,9$, respectively.
Since the faint-end slopes of the Schechter luminosity and mass functions (Eqs.~[\ref{eq:schlum}] and [\ref{eq:schmass}]) are $-0.4(1+\alpha_L)$ and $(1+\alpha_M)$, respectively, the relationship between the two slopes can be derived from $dn = \phi(\Muv) d\Muv = \Phi(M_s)dM_s$ as follows:
\begin{equation}
\beta\,(1+\alpha_M)=-0.4\,(1+\alpha_L). 
\end{equation}
Therefore plugging in $\beta = -0.35$ and solving for $\alpha_M$ results in
\begin{equation}
\alpha_M=0.14+1.14\,\alpha_L, 
\end{equation}
which clarifies the relationship between the two slopes. 
Utilizing this relationship, we are able to recover either $\alpha_M$ or $\alpha_L$ to within $< 0.5$ sigma from one another.  The exact value of $\beta$ must depend on the details of the stellar IMF and star formation histories, and we plan to investigate the evolution of $\beta$ as a function of redshift in more detail in future work. 

%%In this paper, we are not extending our work down to lower redshifts, however, it is possible that the faint galaxies that we find at $z\ge 6$ is related to the faint galaxies that have been identified recently at $z\leq 1$ by \citet{Drory.etal:10} and at $z=2-3$ by \citet{Reddy.etal:10}.  In particular, \citet{Drory.etal:10} found a secondary upturn in GSMF for $M_s<10^{10} \Msun$ galaxies with $\alpha_M=-1.7$ in the COSMOS data. These faint galaxies at lower redshifts could be t%he remains of high-redshift star-forming galaxies we find here.

We see in Figure~\ref{fig:massf} that our simulation results and the observational result of \citet{Gonzalez.etal:10} do not agree well at both faint and bright end, with Gonzalez's GSMF being shallower than our simulation result. 
This can be clearly explained through different $\Muv$ vs. $\log(M_s)$ relationship, as we show in Figure~\ref{fig:muv_ms}. The Gonzalez's relationship (shown as blue long-dashed line; $\log M_s = -0.68 \Muv + {\rm const}$; their Fig.\,1) is much steeper than what our simulation suggests, and for a given value of $\Muv$, they infer a higher $M_s$ for a brighter $\Muv$, and a lower $M_s$ for a fainter $\Muv$. We note that they derived this relationship from $z=4$ data set, and the limited data at $z=6$ does not seem to follow this relationship tightly.  
This difference in the slope of $\Muv$ vs. $\log(M_s)$ relationship is directly reflected in the difference in the GSMF slope. If we translate $\Muv$ into luminosity, then our simulation predicts $L_{\rm UV}\propto M_s^{0.875}$, whereas \citet{Gonzalez.etal:10} obtained  $L_{\rm UV}\propto M_s^{1.7}$ at $z\sim 4$. 
We plan to study the evolution of relationship between $\Muv$, $M_s$ and SFR in our simulations in our subsequent work.

 %%Fig 10

In the next section, we discuss the implications of these steep mass/luminosity functions in our simulations on the reionisation of the Universe.  This is an important subject that is closely related to our current work, as the source of ionising radiation is considered to be coming from these low mass, star forming galaxies at $z\gtrsim 6$.

%%%%%%%%%%%%%%%%%%%%%%%%%%%%%%%%%%%%%%%%%%%%%%%%%%

\section{Implications for Reionisation \& Minimum Halo Mass for Hosting Faint Galaxies}
\label{sec:reion}

\begin{figure}
\begin{center}
\includegraphics[scale=0.43] {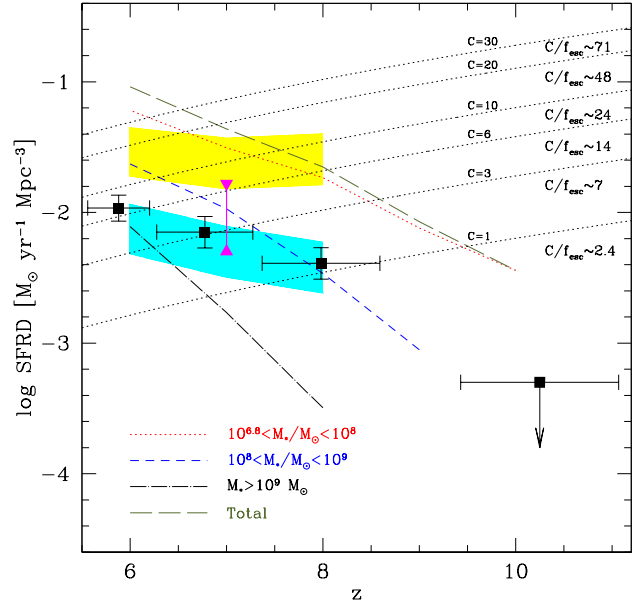}
\caption{
Star formation rate density (SFRD) as a function of redshift.  The result from multiple galaxy stellar mass ranges are shown: $10^{6.8}<M_s/\Msun<10^{8}$ (dotted red), $10^{8}<M_s/\Msun<10^{9}$ (short dashed blue) and $M_s>10^{9} \Msun$ (dashed-dot black).  The total sum of all mass ranges is shown as the dark green dashed line, and the fractional contribution from each mass range can be found in Table~\ref{tbl:sfrd_frac}.  Observational estimates taken from \citet{Bouwens.etal:10b} (black squares) and \citet{Ouchi.etal:09} (magenta triangles).  The yellow and cyan shaded areas represents a range of model estimates presented in \citet{Munoz.etal:10}.  Dotted black contour lines represent minimum SFRD required to keep the IGM ionized as defined by \citet{Madau.etal:99} using varying values for $C$ and a constant value of $f_{esc}=0.42$.
}
\label{fig:sfrd}
\end{center}
\end{figure}

\begin{figure}
\begin{center}
\includegraphics[scale=0.43] {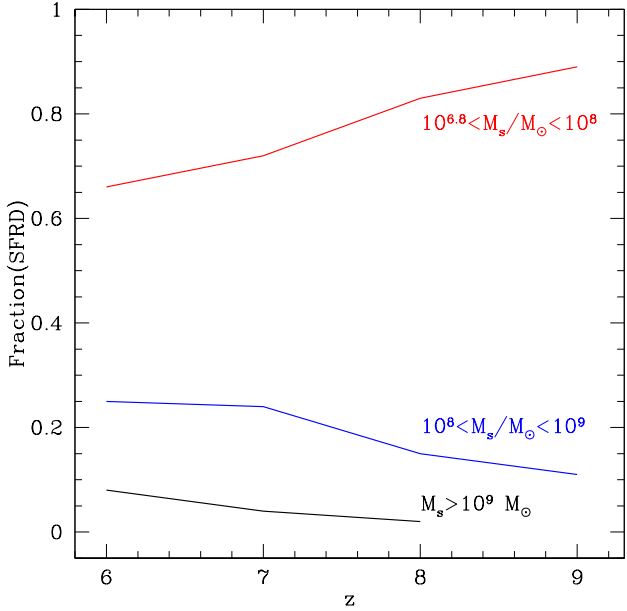}
\caption{Fractional contribution from each galaxy mass range to the total SFRD found in Figure~\ref{fig:sfrd} at $z=6-10$. See Table~\ref{tbl:sfrd_frac} for the numerical values.}
\label{fig:frac_sfrd}
\end{center}
\end{figure}

In order to quantify the contribution of the faint, low-mass galaxies to the reionisation of the Universe, we examine the SFR density (SFRD) from galaxies with different stellar masses at $z\ge 6$. 
Figure~\ref{fig:sfrd} shows the SFRD as a function of redshift broken down into galaxy stellar mass bins of $M_s= 10^{6.8}-10^8 \Msun$ (red), $10^8-10^9 \Msun$ (blue), $>10^9 \Msun$ (black), and the total SFRD (dashed dark green).  
We also show the observational estimates taken from \citet[][black squares]{Bouwens.etal:10b} and \citet[][magenta triangles]{Ouchi.etal:09}. 
The fractional contribution from each mass range is listed in Table~\ref{tbl:sfrd_frac}, and its redshift evolution is summarised in Figure~\ref{fig:frac_sfrd}.

The dotted black lines indicate the SFRD required to maintain reionisation \citep{Madau.etal:99}:
\begin{equation} 
\label{madau}
\dot\rho_\star\approx2\times10^{-3}\left( \frac{C}{f_{esc}}\right ) \left(\frac{1+z}{10}\right)^{3},
\end{equation}
which is dependent on redshift and the ratio of IGM clumping factor ($C$) and escape fraction ($f_{esc}$) of ionizing photons from galaxies.  Although the exact value for $C$ is unknown, a range of values has been estimated from numerical simulations:  $C=30$ \citep{Gnedin.etal:97},  $C=10$ \citep{Iliev.etal:06}, 
and $C=3-6$ \citep{Pawlik.etal:09}. 

In our recent work, \citet{Yajima.etal:10} estimated the values of $f_{esc}$ as a function of halo mass in our simulations, by performing ray-tracing radiative transfer calculations of ionizing photons from star-forming galaxies. 
They found that $f_{esc}$ decreases with increasing halo mass.  In order to obtain a representative value of $f_{esc}$, we compute the average $\langle f_{esc} \rangle$, weighted by the halo number density, using \citet{Sheth.etal:02} halo mass function, and obtain $\langle f_{esc} \rangle = 0.42$.  Here we use this value to calculate $C$ for each assumed value of $C/f_{esc}$. 

%%Fig 11

For comparison, the yellow and cyan shaded areas in Figure~\ref{fig:sfrd} represent a range of model predictions from \citet{Munoz.etal:10}.  They developed a semi-analytic model of star-forming galaxies using extended Press-Schechter theory, and examined the effect of minimum halo mass that can host a galaxy on the UV LF.  They found that the best-fit UV LF to the current observations can be obtained with a minimum halo mass of $M_{\rm h, min}\approx 10^{9.4} \Msun$.  The cyan shade indicates the SFRD derived using this best-fit minimum halo mass and varying $L_{1500}/SFR$.   The same is done considering $M_{\rm h, min} = 10^8 \Msun$, shown by the yellow shade.  By comparison, our simulation is more consistent with the case of  $M_{\rm h, min}\sim 10^8 \Msun$, as confirmed by the direct data in our simulation (see the later discussion in Section~\ref{sec:slope} and Figure~\ref{fig:luv_mh}). In our N400L10 run which resolves the lowest halo mass among the three simulations, we find $M_{\rm h, min}\approx 10^{8.2} \Msun$, corresponding to the virial temperature of $T_{\rm vir} \approx 10^4$\,K. This is expected, because the cooling curve used in our simulation cuts off at around this temperature. 

\begin{table}
\begin{center}
\begin{tabular}{cccc}
\hline
\multicolumn{4}{|c|}{\bf SFRD Fraction} \\
\hline
 $z$ & $10^{6.8}-10^8 \Msun$ & $10^8-10^9\Msun$ &$ > 10^9\Msun$    \\ 
\hline
6 & $0.66$ & $0.25$ & $0.08$  \\
7 & $0.72$ & $0.24$ & $0.04$  \\
8 & $0.83$ & $0.15$ & $0.02$  \\
9 & $0.89$ & $0.11$ & ---  \\
10 & $1.00$ & --- & ---   \\
\hline
\end{tabular}
\caption{Fractional contribution from each galaxy stellar mass range to the total SFRD found in Figure~\ref{fig:sfrd} at $z=6-10$.
The redshift evolution of these values are summarised in Figure~\ref{fig:frac_sfrd}.}
\label{tbl:sfrd_frac}
\end{center}
\end{table} 

\begin{table}
\begin{center}
\begin{tabular}{ccc}
\hline
 $z$ & $\rho_{L}$ & $\rho_{s}$    \\ 
 $ $ & $-18<\Muv < -15$ & $10^{6.8}<M_s < 10^{8.2} \Msun$\\
\hline
6 & $0.64$ & $0.71$  \\
7 & $0.74$ & $0.82$  \\
8 & $0.83$ & $0.89$  \\
\hline
\end{tabular}
\caption{Fractional contribution to the UV Luminosity density ($\rho_L$) and stellar mass density ($\rho_s$) by the faint-end galaxies below current detection limit, calculated by integrating the best-fit Schechter functions.  Limits of integration correspond to the current observational limit ($\Muv=-18$, $M_s=10^{8.2}\Msun$) and to the simulation resolution limit ($\Muv=-15$, $M_s=10^{6.8}\Msun$).  See Sections~\ref{sec:LFs} and \ref{sec:MFs} for details regarding resolution limits. }
\label{tbl:den_frac}
\end{center}
\end{table} 

%%Fig 12

%%Fig 13
\begin{figure}
\begin{center}
\includegraphics[scale=0.43] {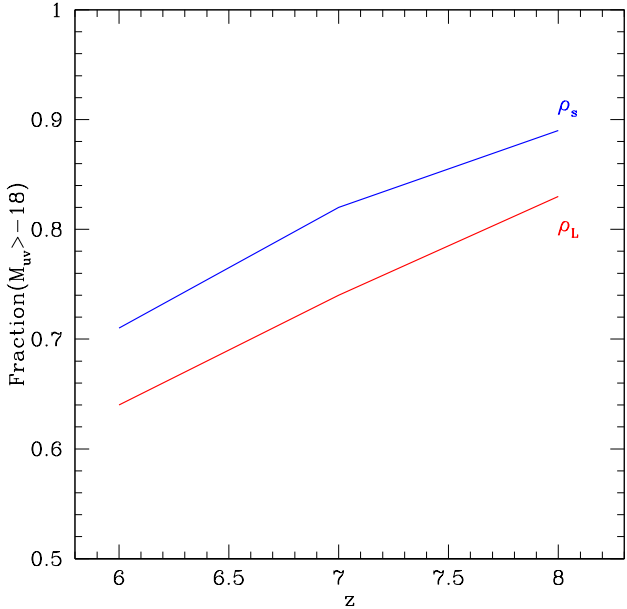}
\caption{Fractional contribution to the total luminosity density ($\rho_L$) and stellar mass density ($\rho_s$) by the galaxies below the current detection limit of $\Muv = -18$. The values are obtained by integrating the best-fit Schechter functions in the range of $-18 < \Muv < -15$ or $M_s = 10^{6.8}-10^8 \Msun$.  See Table~\ref{tbl:den_frac} for numerical values.}
\label{fig:frac_den}
\end{center}
\end{figure}

Using the conditional luminosity function method,  \citet{Trenti.etal:10} found a similar trend of a steepening faint-end slope from $\alpha_L=-1.74$ at $z=6$ to $\alpha_L=-1.99$ at $z=9$, based on the evolution of dark matter halo mass function.  Given these steep faint-end slopes, they argued that galaxies between $\Muv = -18$ and $\Muv = -10$ contribute $\geq 75$\% of the total luminosity density at $z=7$ and produce enough ionizing photons to maintain reionisation assuming $C$/$f_{esc} \lesssim 25$ ($C=5$, $f_{esc}=0.2$).
In our simulations, this fraction is even higher ($>90\%$ for the same $\Muv$ range as above) due to our steeper $\alpha_L$.   However, if we change the integration limit to $-18 < \Muv < -15$ corresponding to our resolution limit, 
we find this fraction to be $74\%$ at $z=7$.  See Table~\ref{tbl:den_frac} for the fractional contribution by the faint-end galaxies to the total UV luminosity density and stellar mass density at $z=6-9$. 

Examining the total SFRD found in our simulations (long-dashed line), we find that, depending on our assumptions of the values for $C$,  galaxies can maintain reionisation as early as $z\sim 9.5$ for $C=1$, and as late as $z\sim6.5$ for $C=30$.  In our simulations, low-mass galaxies with $M_s = 10^{6.8} - 10^8 \Msun$ (red line) and  $M_{UV}>-18$ are the primary contributor to the SFRD at $z\ge 6$ as shown in 
Table~\ref{tbl:sfrd_frac}, and they are capable of achieving reionisation by $z\sim 6$.  Note that our simulations do not include contributions from Pop III stars.

%%%%%%%%%%%%%%%%%%%%%%%%%%%%%%%%%%%%%%%%%
\section{Discussions}
\label{sec:Disc}

\subsection{Numerical Resolution Issues}
\label{sec:Res}

In order to assess the resolution limit of the simulation, one may compare the particle mass and smoothing length to the Jeans mass and length.  Assuming the SF threshold density ($n_{\rm th}^{\rm SF}=0.6$\,cm$^{-3}$) and a temperature range of $500-3000$ K, we find that the Jeans length and mass to be $\lambda_{\rm J} \approx 300 - 700 $\,pc and $M_{\rm J}\approx 10^5-10^7 \Msun$.  Compared to these numbers, the proper gravitational softening length of the N400L10 run at $z=6$ is $\epsilon = 197$\,pc, which is shorter than $\lambda_{\rm J}$ by $100-500$\,pc.  The gas particle mass of the same run ($m_{\rm gas} = 2.65\times 10^5 \Msun$ with $h=0.72$) is lower than $M_{\rm J}$ by a factor of $\approx 1 - 100$.  

However, \citet{Bate&Burkert:97} argued for a mass resolution criteria that requires the local Jeans mass to be less than $M_{\rm res}\approx 2 N_{\rm neigh}$, where $N_{\rm neigh}$ is the number of particles in the SPH kernel.   For the N400L10 run, $N_{\rm neigh} = 33$ was used, resulting in $M_{\rm res} \approx 6.6 \times 10^6\Msun$.  This number is in-between the range of $M_{\rm J}$ given above, and it suggests that our N400L10 simulation cannot resolve the gravitational collapse of gas clouds properly at temperatures below $\sim$2120\,K,  similarly to other cosmological SPH simulations of galaxy formation with similar box sizes and particle counts.  This is one of the reasons why cosmological simulations need to adopt a sub-particle multiphase ISM model and form star particles out of a cold phase gas that co-exist with a hot phase gas with high effective temperature.  Such sub-particle SF models are designed to reproduce the observed \citet{Kennicutt:98} law in a statistical sense \citep[e.g.,][]{Springel:03, Schaye:08, Choi:10a}, even though the simulations cannot resolve the collapse of the cold gas clouds. 

\begin{figure}
\begin{center}
\includegraphics[scale=0.43] {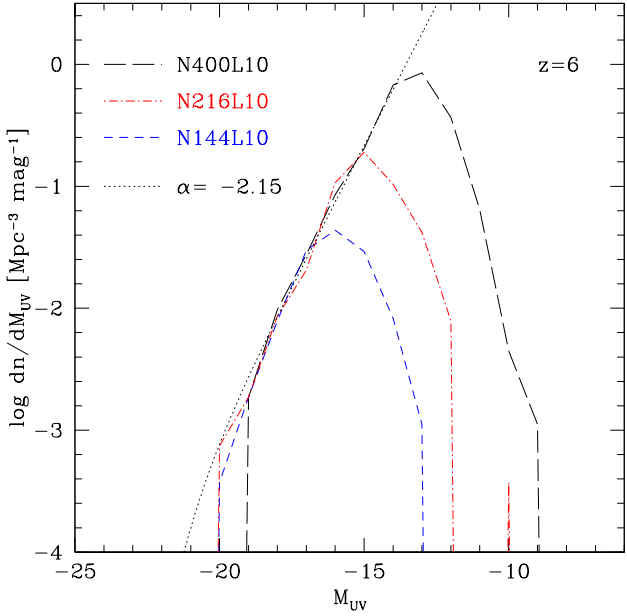}
\caption{
Luminosity functions at $z=6$ from the runs of the same box size ($10 \himpc$) and varied resolutions:  $2\times 144^3$ particles (short-dashed blue line), $2\times 216^3$ particles (dash-dot red line) and $2\times400^3$ particles (long-dashed black line).  The dotted black line represents the faint-end slope of $\alpha=-2.15$.
}
\label{fig:conv}
\end{center}
\end{figure}

Despite of all the discussion above, one would have to perform a convergence study to be really confident that the results are unaffected by the resolution limit.  To show that our results on the faint-end are not resolution dependent, we compare the LFs created from runs with seed particle counts of $2\times400^3$ (long dash black), $2\times216^3$ (dash-dot red) and $2\times144^3$ (short dashed blue) all in  a $10 \himpc$ box size in Figure~\ref{fig:conv}.  Here we see that the faint-end slope from $\Muv = -18$ to $\Muv = -15$ remains largely unchanged in the N400L10 run. 

%%Fig 14

%%Fig. 15
\begin{figure}
\begin{center}
\includegraphics[scale=0.43] {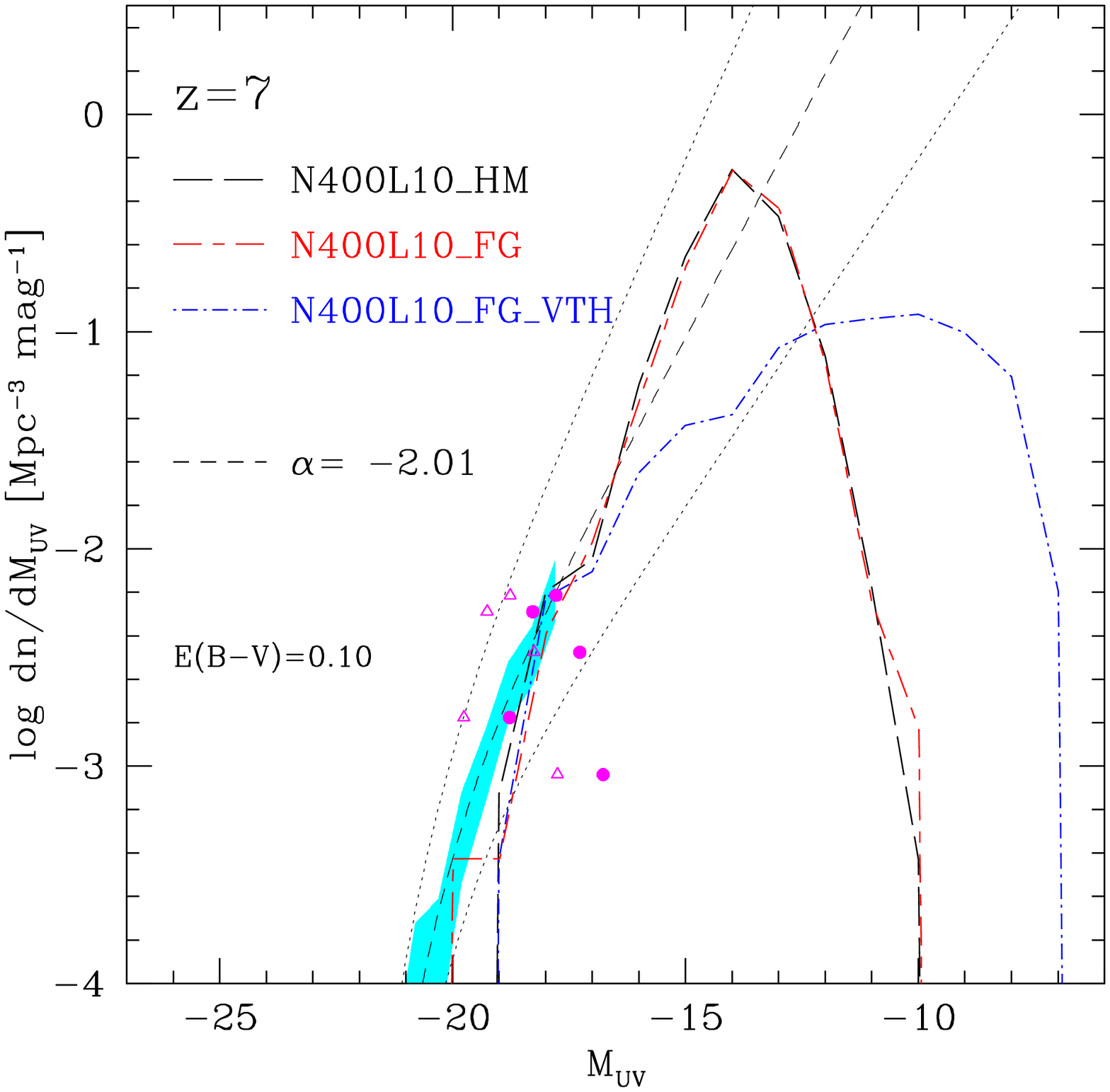}
\caption{
Comparison of $z=7$ LFs using different UVB prescriptions.  The long-dashed black line represents the N400L10 run using a modified \citet{Haardt:96} spectrum \citep{Katz:96, Dave:99}.  The long-short dashed red line is with \citet[][FG-UVB]{Faucher.etal:09} model in the same N400L10 run.  The dash-dot blue line is a N400L10 run using FG-UVB and a variable star formation threshold density (VTH) which has a $(1+z)$ dependency to mimic the possible effect of molecular hydrogen cooling on star formation.  The short-dashed black line (with a faint-end slope of $\alpha=-2.01$) indicates the Schechter function fit to observed data (cyan shade) along with $1\sigma$ fit parameter errors (dotted black lines) by \citet{Bouwens.etal:11a}.  Magenta triangles represent $z=7$ LF with no dust extinction from the \citet{Kuhlen.etal:11} KMT09 simulation, which used an SF model based on the H$_2$ mass. Filled magenta circles approximate the results of applying a dust extinction of $E(B-V)=0.10$.
}
\label{fig:hf_fg_vth}
\end{center}
\end{figure}

%%%%%%%%%%%%%%%%%%%%%%%%%%%%%%%%%%%%%%%%%

\subsection{UVB Radiation and H$_2$ Cooling}
\label{sec:h2}

Among the results presented in this paper, the most striking one is the very steep faint-end slopes in both galaxy mass and luminosity functions at $z\ge 6$, and we find a greater number of low-mass galaxies than what the current observations suggest.  While it is possible that the current observations are missing these faint galaxies at high-redshift and that the upcoming JWST will detect these faint sources, we should also consider the uncertainties and limitations in the treatment of star formation in our numerical simulations. 

In our simulations, stars are allowed to form once the gas density exceeds the threshold density of $n_{\rm th}^{\rm SF}=0.6$\,cm$^{-3}$ \citep[see][for justification of this value]{Nagamine.etal:10}, and star particles are generated according to the SF law matched to the local \citet{Kennicutt:98} law based on the cooling functions with metal line contributions \citep{Choi:10a}.  If the star formation efficiency decreases with increasing redshift in the real Universe, it is possible that our simulations are overproducing stars at high redshift. This could happen, for example, if star formation is controlled primarily by the amount of molecular gas, which is dependent on the gas metallicity.  In the early Universe, ISM and IGM may not be enriched with metals and dust sufficiently yet, which leads to a lower molecular mass densities and lower SFRD \citep{Gnedin.etal:10,Kuhlen.etal:11}.  

To investigate the possible impact of H$_2$ cooling on our result, we experiment with a simple model varying star formation threshold (VTH), in which the value of $n_{\rm th}^{\rm SF}$ is increased by a factor of $(1+z)$ with increasing redshift.  The result of this run is shown by the dash-dot blue line in Figure~\ref{fig:hf_fg_vth}.  As the SF threshold density is increased and the star formation is limited to regions with higher and higher density, the number of low mass galaxies is decreased significantly at $z=7$, showing a much flatter faint-end slope than our fiducial runs. 

In Figure~\ref{fig:hf_fg_vth} we also compare our fiducial and VTH model to the KMT09 simulation by \citet{Kuhlen.etal:11}, which utilizes an SF model based on  H$_2$ mass (magenta triangles).  Although covering only small part of the total LF, we can clearly see that without dust extinction correction, this model overpredicts the observed rest-frame UV LF at $z=7$, similarly to our simulations.  When we apply a dust extinction effect with $E(B-V)=0.10$ approximately to the Kuhlen et al.'s data (a shift of $\sim$1 mag similarly to our simulations), their result (filled magenta circles) agree with observations very well just like our result. This suggests  that the simulations utilizing a SF model based on H$_2$ mass produce a very similar LF to that presented here, except the sharp turnover at $\Muv \sim -18$ seen for the Kuhlen et al.'s data. 
We are currently in the process of implementing a similar SF model based on H$_2$ mass in our simulations, and we will report the results in a subsequent paper (Thompson, et al., in preparation).

%%Fig. 16
\begin{figure}
\begin{center}
\includegraphics[scale=0.43] {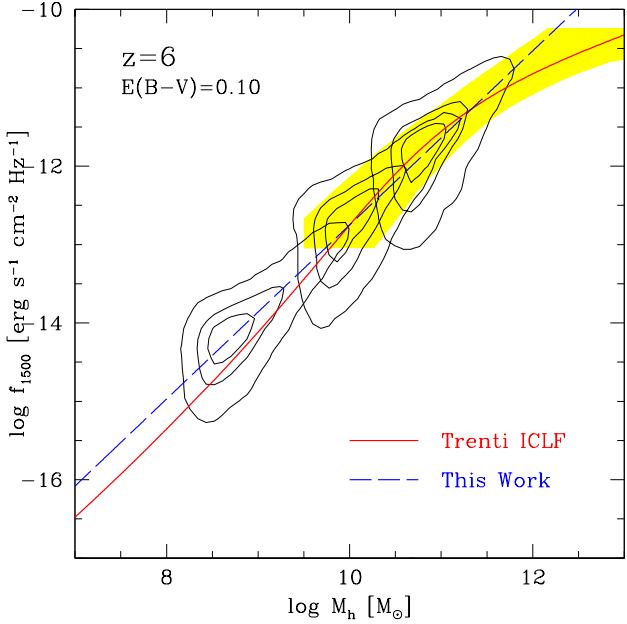}
\caption{
Contour plot of absolute UV flux corresponding to AB magnitude at $10 {\rm pc}$ plotted against dark matter halo mass for each run N400L10 (left), N400L34 (middle) and N600L100 (right). The luminosity for each halo is a summation of all galaxies found in the halo.  Blue dashed line show a least square fit to the data.  Solid red line represents the results of a semi analytic model developed by \citet{Trenti.etal:10}.  Yellow shaded region shows a range of predictions based on semi-analytic calculations found in \citet{Lee.etal:10}. 
}
\label{fig:luv_mh}
\end{center}
\end{figure}

%%%%%%%%%%%%%%%%%%%%%%%%%%%%%%%%%%%%%

\subsection{Origin of Steep Faint End Slope}
\label{sec:slope}

In order to compare our result with some of the semi-analytic models based on the halo occupation model, in Figure \ref{fig:luv_mh} we examine the relationship between the rest-frame, monochromatic flux $f_{1500}$ at 1500 \AA\ and the halo mass $M_h$ for all the halos in our simulations at $z=6$.  
The dashed blue line is a least-square fit to all the data points shown by the contour, the solid red line is the result of semi-analytic modeling by \citet{Trenti.etal:10}, and the yellow shade is another similar model result by \citet{Lee.etal:10}.   
Trenti et al. calibrated their model to the observed LF at $z=6$, and Lee et al. to the observed LF and correlation functions at $z=4,5,6$, with both assuming single galaxy occupation per halo. 
Lee et al.'s model implements a mass and luminosity threshold for galaxies, thus the yellow shade does not cover the entire range of our simulation.  

We see good agreement between these model results at $M_h \gtrsim 10^{10} \Msun$. 
Below $M_h=10^{10} \Msun$ our relationship begins to deviate from Trenti's model, being higher by $\sim0.5$ dex.  This result implies that, in our simulations the low mass halos with $<10^{10} \Msun$ are more efficient at producing stars than in the model of Trenti et al., making our faint-end slope steeper than theirs and Bouwens et al.'s. 

\begin{figure}
\begin{center}
\includegraphics[scale=0.43] {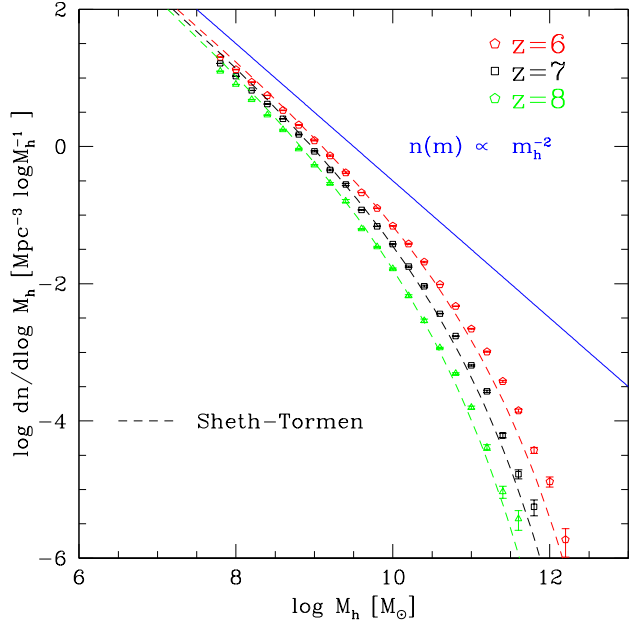}
\caption{
Composite dark matter halo mass function (HMF) at $z=6$ (red), $z=7$ (black) and $z=8$ (green) assembled at each redshift from the N600L100, N400L34 and N400L10 runs.  The dashed line at each redshift represents predicted halo mass function based on theoretical work found in \citet{Sheth.etal:02}.  Blue solid line indicates a slope$=-2.00$  offset for comparison.
}
\label{fig:hmf}
\end{center}
\end{figure}

To further validate our simulation results, in Figure \ref{fig:hmf} we compare the dark matter halo mass function (HMF) in our simulations (data points) to the analytic model of \citet[][dashed line]{Sheth.etal:02}. 
The HMF was assembled from our three runs in the same manner as the LF and GSMF. 
Here we find very good agreement with the model, indicating that we are producing proper number of halos in the mass range of $10^8 \lesssim M_h \lesssim 10^{12} \Msun$. 

Since we are producing correct number of halos, the above results suggest that our steeper than observed faint-end slope can be attributed to the fact that the low mass halos are producing stars more efficiently, rather than too many galaxies in a particular $\Muv$ bin.   This would have the effect of steepening the faint-end slope, since the deviation occurs primarily in the N400L10 run which controls the faint-end of the LF.  Further work must be done to explain the exact baryonic physics of this effect. 

%%Fig 17

%%%%%%%%%%%%%%%%%%%%%%%%%%%%%%%%%%%%%%%%%

\section{Conclusions}
\label{sec:summary}

Using cosmological SPH simulations, we examined the colors of high-redshift galaxies, quantified the evolution of luminosity and mass functions at $z=6-9$, compared our results to the latest WFC3 observational results, and examined the implications for the reionisation of the Universe.   Following are the main conclusions of the present work:

\begin{itemize}
\item  We find good agreement with observations in color-color space at $z=6, 7$, and 8.  Most of the simulated galaxies with $E(B-V)=0.0 - 0.30$ are consistent with the color selection criteria with a relatively tight distribution. This suggests that the galaxies selected by the current color selection criteria could have a wide range of extinction values with $E(B-V)=0.0 - 0.30$.  We also find that the scatter in the observed data on the color-color plane is more dominated by the redshift scatter and photometric errors, rather than the variance in extinction. 

\vspace{0.2cm}
\item  The rest-frame UV LF of simulated galaxies agree well with the current WFC3 observations when we assume uniform dust extinctions of $E(B-V) = 0.10$ for $z=6,7$ and $E(B-V) = 0.075$ for $z=8$. These extinction values are consistent with those obtained by \citet{Schaerer.deBarros:10}, who performed SED fitting including nebular emission lines. See Section~\ref{sec:LFs} for more details on the extinction treatment. 

\vspace{0.2cm}
\item We performed least $\chi^{2}$ fits to simulation LFs using the three parameter Schechter luminosity and mass functions.  The best-fit Schechter parameters are summarised in Tables~\ref{tbl:sch} and \ref{tbl:schm}.
Each of the three Schechter parameters, ($\alpha_L$, $\Muv^*$, and $\phi_L^*$) or ($\alpha_M$, $M_s^*$, and $\phi_M^*$), are evolving with redshift.  The faint-end slope is steeper than the current observational estimates with $\alpha<-2$, and becomes steeper with $|\alpha_L| \propto (1+z)^{0.59}$ and $|\alpha_M| \propto (1+z)^{0.65}$. 
The characteristic mass $M_s^{*}$ decreases towards higher redshift with $M_s^{*} \propto (1+z)^{-0.20}$. 
The characteristic magnitude $\Muv^{*}$ does not evolve very much, but it does get dimmer from $z=6$ to $z=7$ by about 0.3 magnitude.  The normalization $\phi^*$ decreases towards higher redshift as expected in the hierarchical structure formation model, with $\phi_L^* \propto (1+z)^{-0.42}$ and $\phi_M^* \propto (1+z)^{-0.45}$. 

\vspace{0.2cm}
\item We decomposed the total SFRD into three separate contributions from different galaxy mass ranges of $M_s = 10^{6.8}-10^8 \Msun, 10^8-10^9 \Msun$ and $>10^9 \Msun$, in a similar fashion to the work by \citet{Choi:11b}.  We find that galaxies with $M_s = 10^{6.8} - 10^8 \Msun$ are the primary contributor to the total SFRD at $z\ge 6$ as shown in Figs.~\ref{fig:sfrd}, \ref{fig:frac_sfrd}, and Table~\ref{tbl:sfrd_frac}, 
and therefore to the ionizing photon budget as well. Our simulation suggests that these faint galaxies can reionise the Universe by $z=6$ as long as the clumping factor is $C<30$.  This conclusion is consistent with that of our earlier direct radiative transfer calculations by \citet{Yajima.etal:10}.  

\vspace{0.2cm}
\item We find that the contribution from low-mass galaxies with $M_s = 10^{6.8}-10^8 \Msun$ to the total SFRD (and hence the ionizing photon budget) decreases with decreasing redshift, from $89\%$ at $z=9$ to $66\%$ at $z=6$, which is still dominant.  
This decreasing trend can also be seen in the contribution from faint-end galaxies ($-18 < \Muv < -15$) to the total UV luminosity density $\rho_L$ and stellar mass density $\rho_s$ (see Table~\ref{tbl:den_frac} and Figure~\ref{fig:frac_den}).  
Whereas the fractional contribution from galaxies with $M_s = 10^8-10^9 \Msun$ and $>10^9 \Msun$ are increasing gradually during the same epoch with much lower fraction (see Table~\ref{tbl:sfrd_frac} and Figure~\ref{fig:frac_sfrd}).  
The fractional contribution from galaxies with $M_s >10^9 \Msun$ is particularly small with less than 10\% at all redshifts of $z=6-9$. 

\end{itemize}

In summary, the steep faint-end slopes of high-redshift galaxies at $z\ge 6$ seems to be a generic prediction of $\Lambda$CDM model, {\em if} one calibrates the SF model to match the locally observed \citet{Kennicutt:98} law or to the lower redshift LF results. 
In addition to the results of our simulations, the two simple semianalytic models obtain similarly steep faint-end slopes \citep{Munoz.etal:10, Trenti.etal:10}, as well as the more detailed semianalytic model of galaxy formation by \citet{LoFaro.etal:09}.  The existence of these low-mass galaxies below the current detection limit can be tested soon by the JWST, and we will continue to investigate the possibility of decreasing SF efficiency towards high-redshift by implementing more sophisticated physical processes (e.g., metallicity dependent SF model based on H$_2$ masses)  in our future simulations.  
Given the significant contribution by the faint-end galaxies to the overall budget of SFRD, ionizing photons, luminosity and stellar mass densities, it is critical to determine the number density and the nature of this population more accurately.

\section*{Acknowledgments}

We are grateful to V. Springel for allowing us to use the original version of GADGET-3 code, on which the \citet{Choi:10a, Choi:11b} simulations are based.  We would also like to thank M. Trenti for communications regarding comparisons made in Figure \ref{fig:luv_mh}. 
This work is supported in part by the NSF grant AST-0807491, NASA grant HST-AR-12143-01-A, National Aeronautics and Space Administration under Grant/Cooperative Agreement No. NNX08AE57A issued by the Nevada NASA EPSCoR program, and the President's Infrastructure Award from UNLV. This research is also supported by the NSF through the TeraGrid resources provided by the Texas Advanced Computing Center. Some numerical simulations and analyses have also been performed on the UNLV Cosmology Cluster.

%\bibliographystyle{mn2e}
%\bibliography{test}
%\bibliography{jjref}

\begin{thebibliography}{}

\bibitem[\protect\citeauthoryear{{Andrae}}{{Andrae}}{2010}]{Andrae:10}
{Andrae} R.,  2010, ArXiv e-prints

\bibitem[\protect\citeauthoryear{{Bate} \& {Burkert}}{{Bate} \&
  {Burkert}}{1997}]{Bate&Burkert:97}
{Bate} M.~R.,  {Burkert} A.,  1997, \mnras, 288, 1060

\bibitem[\protect\citeauthoryear{{Beckwith}, {Stiavelli}, {Koekemoer},
  {Caldwell}, {Ferguson}, {Hook}, {Lucas}, {Bergeron} et~al.,}{{Beckwith}
  et~al.}{2006}]{Beckwith.etal:06}
{Beckwith} S.~V.~W.,  {Stiavelli} M.,  {Koekemoer} A.~M.,  {Caldwell} J.~A.~R.,
   {Ferguson} H.~C.,  {Hook} R.,  {Lucas} R.~A.,  {Bergeron} L.~E.,    et~al.,
  2006, \aj, 132, 1729

\bibitem[\protect\citeauthoryear{{Bouwens}, {Illingworth}, {Blakeslee} \&
  {Franx}}{{Bouwens} et~al.}{2006}]{Bouwens.etal:06}
{Bouwens} R.~J.,  {Illingworth} G.~D.,  {Blakeslee} J.~P.,    {Franx} M.,
  2006, \apj, 653, 53

\bibitem[\protect\citeauthoryear{{Bouwens}, {Illingworth}, {Franx}, {Chary},
  {Meurer}, {Conselice}, {Ford}, {Giavalisco} et~al.,}{{Bouwens}
  et~al.}{2009}]{Bouwens.etal:09}
{Bouwens} R.~J.,  {Illingworth} G.~D.,  {Franx} M.,  {Chary} R.,  {Meurer}
  G.~R.,  {Conselice} C.~J.,  {Ford} H.,  {Giavalisco} M.,    et~al., 2009,
  \apj, 705, 936

\bibitem[\protect\citeauthoryear{{Bouwens}, {Illingworth}, {Franx} \&
  {Ford}}{{Bouwens} et~al.}{2008}]{Bouwens.etal:08}
{Bouwens} R.~J.,  {Illingworth} G.~D.,  {Franx} M.,    {Ford} H.,  2008, \apj,
  686, 230

\bibitem[\protect\citeauthoryear{{Bouwens}, {Illingworth}, {Oesch}, {Labbe},
  {Trenti}, {van Dokkum}, {Franx}, {Stiavelli} et~al.,}{{Bouwens}
  et~al.}{2010}]{Bouwens.etal:10b}
{Bouwens} R.~J.,  {Illingworth} G.~D.,  {Oesch} P.~A.,  {Labbe} I.,  {Trenti}
  M.,  {van Dokkum} P.,  {Franx} M.,  {Stiavelli} M.,    et~al., 2010, ArXiv
  e-prints

\bibitem[\protect\citeauthoryear{{Bouwens}, {Illingworth}, {Oesch},
  {Stiavelli}, {van Dokkum}, {Trenti}, {Magee}, {Labb{\'e}} et~al.,}{{Bouwens}
  et~al.}{2010}]{Bouwens.etal:10a}
{Bouwens} R.~J.,  {Illingworth} G.~D.,  {Oesch} P.~A.,  {Stiavelli} M.,  {van
  Dokkum} P.,  {Trenti} M.,  {Magee} D.,  {Labb{\'e}} I.,    et~al., 2010,
  \apjl, 709, L133

\bibitem[\protect\citeauthoryear{{Bouwens}, {Illingworth}, {Oesch}, {Trenti},
  {Labbe}, {Franx}, {Stiavelli}, {Carollo}, {van Dokkum} \& {Magee}}{{Bouwens}
  et~al.}{2011}]{Bouwens.etal:11a}
{Bouwens} R.~J.,  {Illingworth} G.~D.,  {Oesch} P.~A.,  {Trenti} M.,  {Labbe}
  I.,  {Franx} M.,  {Stiavelli} M.,  {Carollo} C.~M.,  {van Dokkum} P.,
  {Magee} D.,  2011, ArXiv e-prints

\bibitem[\protect\citeauthoryear{{Bouwens}, {Illingworth}, {Thompson},
  {Blakeslee}, {Dickinson}, {Broadhurst}, {Eisenstein}, {Fan}
  et~al.,}{{Bouwens} et~al.}{2004}]{Bouwens.etal:04}
{Bouwens} R.~J.,  {Illingworth} G.~D.,  {Thompson} R.~I.,  {Blakeslee} J.~P.,
  {Dickinson} M.~E.,  {Broadhurst} T.~J.,  {Eisenstein} D.~J.,  {Fan} X.,
  et~al., 2004, \apjl, 606, L25

\bibitem[\protect\citeauthoryear{{Bruzual} \& {Charlot}}{{Bruzual} \&
  {Charlot}}{2003}]{BC03}
{Bruzual} G.,  {Charlot} S.,  2003, \mnras, 344, 1000

\bibitem[\protect\citeauthoryear{{Bunker}, {Stanway}, {Ellis} \&
  {McMahon}}{{Bunker} et~al.}{2004}]{Bunker.etal:04}
{Bunker} A.~J.,  {Stanway} E.~R.,  {Ellis} R.~S.,    {McMahon} R.~G.,  2004,
  \mnras, 355, 374

\bibitem[\protect\citeauthoryear{{Calzetti}}{{Calzetti}}{1997}]{Calzetti:97}
{Calzetti} D.,  1997, \aj, 113, 162

\bibitem[\protect\citeauthoryear{{Castellano}, {Fontana}, {Boutsia}, {Grazian},
  {Pentericci}, {Bouwens}, {Dickinson}, {Giavalisco} et~al.,}{{Castellano}
  et~al.}{2010}]{Castellano.etal:10}
{Castellano} M.,  {Fontana} A.,  {Boutsia} K.,  {Grazian} A.,  {Pentericci} L.,
   {Bouwens} R.,  {Dickinson} M.,  {Giavalisco} M.,    et~al., 2010, \aap, 511,
  A20

\bibitem[\protect\citeauthoryear{{Choi} \& {Nagamine}}{{Choi} \&
  {Nagamine}}{2009}]{Choi:09}
{Choi} J.,  {Nagamine} K.,  2009, \mnras, 395, 1776

\bibitem[\protect\citeauthoryear{{Choi} \& {Nagamine}}{{Choi} \&
  {Nagamine}}{2010}]{Choi:10a}
{Choi} J.-H.,  {Nagamine} K.,  2010, \mnras, 407, 1464

\bibitem[\protect\citeauthoryear{{Choi} \& {Nagamine}}{{Choi} \&
  {Nagamine}}{2011}]{Choi:11b}
{Choi} J.-H.,  {Nagamine} K.,  2011, ArXiv e-prints

\bibitem[\protect\citeauthoryear{{Dav{\'e}}, {Hernquist}, {Katz} \&
  {Weinberg}}{{Dav{\'e}} et~al.}{1999}]{Dave:99}
{Dav{\'e}} R.,  {Hernquist} L.,  {Katz} N.,    {Weinberg} D.~H.,  1999, \apj,
  511, 521

\bibitem[\protect\citeauthoryear{{Dickinson}, {Stern}, {Giavalisco},
  {Ferguson}, {Tsvetanov}, {Chornock}, {Cristiani}, {Dawson}
  et~al.,}{{Dickinson} et~al.}{2004}]{Dickinson:04}
{Dickinson} M.,  {Stern} D.,  {Giavalisco} M.,  {Ferguson} H.~C.,  {Tsvetanov}
  Z.,  {Chornock} R.,  {Cristiani} S.,  {Dawson} S.,    et~al., 2004, \apjl,
  600, L99

\bibitem[\protect\citeauthoryear{{Drory}, {Bundy}, {Leauthaud}, {Scoville},
  {Capak}, {Ilbert}, {Kartaltepe} \& {Kneib}}{{Drory}
  et~al.}{2009}]{Drory.etal:10}
{Drory} N.,  {Bundy} K.,  {Leauthaud} A.,  {Scoville} N.,  {Capak} P.,
  {Ilbert} O.,  {Kartaltepe} J.~S.,    {Kneib} J.~P.~o.,  2009, \apj, 707, 1595

\bibitem[\protect\citeauthoryear{{Efstathiou}}{{Efstathiou}}{1992}]{Efstathiou:92}
{Efstathiou} G.,  1992, \mnras, 256, 43P

\bibitem[\protect\citeauthoryear{{Eisenstein} \& {Hu}}{{Eisenstein} \&
  {Hu}}{1999}]{Eisenstein&Hu:99}
{Eisenstein} D.~J.,  {Hu} W.,  1999, \apj, 511, 5

\bibitem[\protect\citeauthoryear{{Faucher-Gigu{\`e}re}, {Lidz}, {Zaldarriaga}
  \& {Hernquist}}{{Faucher-Gigu{\`e}re} et~al.}{2009}]{Faucher.etal:09}
{Faucher-Gigu{\`e}re} C.-A.,  {Lidz} A.,  {Zaldarriaga} M.,    {Hernquist} L.,
  2009, \apj, 703, 1416

\bibitem[\protect\citeauthoryear{{Finlator}, {Dav{\'e}}, {Papovich} \&
  {Hernquist}}{{Finlator} et~al.}{2006}]{Finlator.etal:06}
{Finlator} K.,  {Dav{\'e}} R.,  {Papovich} C.,    {Hernquist} L.,  2006, \apj,
  639, 672

\bibitem[\protect\citeauthoryear{{Gnedin} \& {Kravtsov}}{{Gnedin} \&
  {Kravtsov}}{2010}]{Gnedin.etal:10}
{Gnedin} N.~Y.,  {Kravtsov} A.~V.,  2010, \apj, 714, 287

\bibitem[\protect\citeauthoryear{{Gnedin} \& {Ostriker}}{{Gnedin} \&
  {Ostriker}}{1997}]{Gnedin.etal:97}
{Gnedin} N.~Y.,  {Ostriker} J.~P.,  1997, \apj, 486, 581

\bibitem[\protect\citeauthoryear{{Gonzalez}, {Labbe}, {Bouwens}, {Illingworth},
  {Franx} \& {Kriek}}{{Gonzalez} et~al.}{2010}]{Gonzalez.etal:10}
{Gonzalez} V.,  {Labbe} I.,  {Bouwens} R.,  {Illingworth} G.,  {Franx} M.,
  {Kriek} M.,  2010, ArXiv e-prints

\bibitem[\protect\citeauthoryear{{Haardt} \& {Madau}}{{Haardt} \&
  {Madau}}{1996}]{Haardt:96}
{Haardt} F.,  {Madau} P.,  1996, \apj, 461, 20

\bibitem[\protect\citeauthoryear{{Iliev}, {Ciardi}, {Alvarez}, {Maselli},
  {Ferrara}, {Gnedin}, {Mellema}, {Nakamoto} et~al.,}{{Iliev}
  et~al.}{2006}]{Iliev.etal:06}
{Iliev} I.~T.,  {Ciardi} B.,  {Alvarez} M.~A.,  {Maselli} A.,  {Ferrara} A.,
  {Gnedin} N.~Y.,  {Mellema} G.,  {Nakamoto} T.,    et~al., 2006, \mnras, 371,
  1057

\bibitem[\protect\citeauthoryear{{Katz}, {Weinberg} \& {Hernquist}}{{Katz}
  et~al.}{1996}]{Katz:96}
{Katz} N.,  {Weinberg} D.~H.,    {Hernquist} L.,  1996, \apjs, 105, 19

\bibitem[\protect\citeauthoryear{{Kennicutt}
  Jr.}{{Kennicutt}}{1998}]{Kennicutt:98}
{Kennicutt} Jr. R.~C.,  1998, \araa, 36, 189

\bibitem[\protect\citeauthoryear{{Komatsu}, {Smith}, {Dunkley}, {Bennett},
  {Gold}, {Hinshaw}, {Jarosik}, {Larson} et~al.,}{{Komatsu}
  et~al.}{2010}]{Komatsu.etal:10}
{Komatsu} E.,  {Smith} K.~M.,  {Dunkley} J.,  {Bennett} C.~L.,  {Gold} B.,
  {Hinshaw} G.,  {Jarosik} N.,  {Larson} D.,    et~al., 2010, ArXiv e-prints

\bibitem[\protect\citeauthoryear{{Kuhlen}, {Krumholz}, {Madau}, {Smith} \&
  {Wise}}{{Kuhlen} et~al.}{2011}]{Kuhlen.etal:11}
{Kuhlen} M.,  {Krumholz} M.,  {Madau} P.,  {Smith} B.,    {Wise} J.,  2011,
  ArXiv e-prints

\bibitem[\protect\citeauthoryear{{Lee}, {Giavalisco}, {Conroy}, {Wechsler},
  {Ferguson}, {Somerville}, {Dickinson} \& {Urry}}{{Lee}
  et~al.}{2009}]{Lee.etal:10}
{Lee} K.-S.,  {Giavalisco} M.,  {Conroy} C.,  {Wechsler} R.~H.,  {Ferguson}
  H.~C.,  {Somerville} R.~S.,  {Dickinson} M.~E.,    {Urry} C.~M.,  2009, \apj,
  695, 368

\bibitem[\protect\citeauthoryear{{Lo Faro}, {Monaco}, {Vanzella}, {Fontanot},
  {Silva} \& {Cristiani}}{{Lo Faro} et~al.}{2009}]{LoFaro.etal:09}
{Lo Faro} B.,  {Monaco} P.,  {Vanzella} E.,  {Fontanot} F.,  {Silva} L.,
  {Cristiani} S.,  2009, \mnras, 399, 827

\bibitem[\protect\citeauthoryear{{Madau}}{{Madau}}{1995}]{Madau:95}
{Madau} P.,  1995, \apj, 441, 18

\bibitem[\protect\citeauthoryear{{Madau}, {Haardt} \& {Rees}}{{Madau}
  et~al.}{1999}]{Madau.etal:99}
{Madau} P.,  {Haardt} F.,    {Rees} M.~J.,  1999, \apj, 514, 648

\bibitem[\protect\citeauthoryear{{Malhotra}, {Rhoads}, {Pirzkal}, {Haiman},
  {Xu}, {Daddi}, {Yan}, {Bergeron} et~al.,}{{Malhotra}
  et~al.}{2005}]{Malhotra:05}
{Malhotra} S.,  {Rhoads} J.~E.,  {Pirzkal} N.,  {Haiman} Z.,  {Xu} C.,  {Daddi}
  E.,  {Yan} H.,  {Bergeron} L.~E.,    et~al., 2005, \apj, 626, 666

\bibitem[\protect\citeauthoryear{{Mannucci}, {Buttery}, {Maiolino}, {Marconi}
  \& {Pozzetti}}{{Mannucci} et~al.}{2007}]{Mannucci.etal:07}
{Mannucci} F.,  {Buttery} H.,  {Maiolino} R.,  {Marconi} A.,    {Pozzetti} L.,
  2007, \aap, 461, 423

\bibitem[\protect\citeauthoryear{{McLure}, {Dunlop}, {Cirasuolo}, {Koekemoer},
  {Sabbi}, {Stark}, {Targett} \& {Ellis}}{{McLure}
  et~al.}{2010}]{McLure.etal:10}
{McLure} R.~J.,  {Dunlop} J.~S.,  {Cirasuolo} M.,  {Koekemoer} A.~M.,  {Sabbi}
  E.,  {Stark} D.~P.,  {Targett} T.~A.,    {Ellis} R.~S.,  2010, \mnras, 403, 960

\bibitem[\protect\citeauthoryear{{Mu{\~n}oz} \& {Loeb}}{{Mu{\~n}oz} \&
  {Loeb}}{2010}]{Munoz.etal:10}
{Mu{\~n}oz} J.~A.,  {Loeb} A.,  2010, ArXiv e-prints

\bibitem[\protect\citeauthoryear{{Nagamine}, {Cen}, {Hernquist}, {Ostriker} \&
  {Springel}}{{Nagamine} et~al.}{2005a}]{Nagamine:05a}
{Nagamine} K.,  {Cen} R.,  {Hernquist} L.,  {Ostriker} J.~P.,    {Springel} V.,
   2005a, \apj, 627, 608

\bibitem[\protect\citeauthoryear{{Nagamine}, {Cen}, {Hernquist}, {Ostriker} \&
  {Springel}}{{Nagamine} et~al.}{2005b}]{Nagamine:05b}
{Nagamine} K.,  {Cen} R.,  {Hernquist} L.,  {Ostriker} J.~P.,    {Springel} V.,
   2005b, \apj, 618, 23

\bibitem[\protect\citeauthoryear{{Nagamine}, {Choi} \& {Yajima}}{{Nagamine}
  et~al.}{2010}]{Nagamine.etal:10}
{Nagamine} K.,  {Choi} J.-H.,    {Yajima} H.,  2010, \apjl, 725, L219

\bibitem[\protect\citeauthoryear{{Nagamine}, {Springel}, {Hernquist} \&
  {Machacek}}{{Nagamine} et~al.}{2004}]{Nagamine:04}
{Nagamine} K.,  {Springel} V.,  {Hernquist} L.,    {Machacek} M.,  2004,
  \mnras, 350, 385

\bibitem[\protect\citeauthoryear{{Night}, {Nagamine}, {Springel} \&
  {Hernquist}}{{Night} et~al.}{2006}]{Night.etal:06}
{Night} C.,  {Nagamine} K.,  {Springel} V.,    {Hernquist} L.,  2006, \mnras,
  366, 705

\bibitem[\protect\citeauthoryear{{Oesch}, {Bouwens}, {Illingworth}, {Carollo},
  {Franx}, {Labb{\'e}}, {Magee}, {Stiavelli} et~al.,}{{Oesch}
  et~al.}{2010}]{Oesch.etal:10b}
{Oesch} P.~A.,  {Bouwens} R.~J.,  {Illingworth} G.~D.,  {Carollo} C.~M.,
  {Franx} M.,  {Labb{\'e}} I.,  {Magee} D.,  {Stiavelli} M.,    et~al., 2010,
  \apjl, 709, L16

\bibitem[\protect\citeauthoryear{{Oesch}, {Carollo}, {Stiavelli}, {Trenti},
  {Bergeron}, {Koekemoer}, {Lucas}, {Pavlovsky} et~al.,}{{Oesch}
  et~al.}{2009}]{Oesch.etal:09}
{Oesch} P.~A.,  {Carollo} C.~M.,  {Stiavelli} M.,  {Trenti} M.,  {Bergeron}
  L.~E.,  {Koekemoer} A.~M.,  {Lucas} R.~A.,  {Pavlovsky} C.~M.,    et~al.,
  2009, \apj, 690, 1350

\bibitem[\protect\citeauthoryear{{Ouchi}, {Mobasher}, {Shimasaku}, {Ferguson},
  {Fall}, {Ono}, {Kashikawa}, {Morokuma}, {Nakajima}, {Okamura}, {Dickinson},
  {Giavalisco} \& {Ohta}}{{Ouchi} et~al.}{2009}]{Ouchi.etal:09}
{Ouchi} M.,  {Mobasher} B.,  {Shimasaku} K.,  {Ferguson} H.~C.,  {Fall} S.~M.,
  {Ono} Y.,  {Kashikawa} N.,  {Morokuma} T.,  {Nakajima} K.,  {Okamura} S.,
  {Dickinson} M.,  {Giavalisco} M.,    {Ohta} K.,  2009, \apj, 706, 1136

\bibitem[\protect\citeauthoryear{{Pawlik}, {Schaye} \& {van
  Scherpenzeel}}{{Pawlik} et~al.}{2009}]{Pawlik.etal:09}
{Pawlik} A.~H.,  {Schaye} J.,    {van Scherpenzeel} E.,  2009, \mnras, 394,
  1812

\bibitem[\protect\citeauthoryear{{Reddy} \& {Steidel}}{{Reddy} \&
  {Steidel}}{2009}]{Reddy.etal:10}
{Reddy} N.~A.,  {Steidel} C.~C.,  2009, \apj, 692, 778

\bibitem[\protect\citeauthoryear{{Richard}, {Pell{\'o}}, {Schaerer}, {Le
  Borgne} \& {Kneib}}{{Richard} et~al.}{2006}]{Richard.etal:06}
{Richard} J.,  {Pell{\'o}} R.,  {Schaerer} D.,  {Le Borgne} J.,    {Kneib} J.,
  2006, \aap, 456, 861

\bibitem[\protect\citeauthoryear{{Robertson}, {Li}, {Cox}, {Hernquist} \&
  {Hopkins}}{{Robertson} et~al.}{2007}]{Robertson.etal:07}
{Robertson} B.,  {Li} Y.,  {Cox} T.~J.,  {Hernquist} L.,    {Hopkins} P.~F.,
  2007, \apj, 667, 60

\bibitem[\protect\citeauthoryear{{Schaerer} \& {de Barros}}{{Schaerer} \& {de
  Barros}}{2010}]{Schaerer.deBarros:10}
{Schaerer} D.,  {de Barros} S.,  2010, \aap, 515, A73+

\bibitem[\protect\citeauthoryear{{Schaye} \& {Dalla Vecchia}}{{Schaye} \&
  {Dalla Vecchia}}{2008}]{Schaye:08}
{Schaye} J.,  {Dalla Vecchia} C.,  2008, \mnras, 383, 1210

\bibitem[\protect\citeauthoryear{{Schechter}}{{Schechter}}{1976}]{Schechter:76}
{Schechter} P.,  1976, \apj, 203, 297

\bibitem[\protect\citeauthoryear{{Sheth} \& {Tormen}}{{Sheth} \&
  {Tormen}}{2002}]{Sheth.etal:02}
{Sheth} R.~K.,  {Tormen} G.,  2002, \mnras, 329, 61

\bibitem[\protect\citeauthoryear{{Springel}}{{Springel}}{2005}]{Springel:05}
{Springel} V.,  2005, \mnras, 364, 1105

\bibitem[\protect\citeauthoryear{{Springel} \& {Hernquist}}{{Springel} \&
  {Hernquist}}{2003}]{Springel:03}
{Springel} V.,  {Hernquist} L.,  2003, \mnras, 339, 289

\bibitem[\protect\citeauthoryear{{Springel}, {White}, {Tormen} \&
  {Kauffmann}}{{Springel} et~al.}{2001}]{Springel:01}
{Springel} V.,  {White} S.~D.~M.,  {Tormen} G.,    {Kauffmann} G.,  2001,
  \mnras, 328, 726

\bibitem[\protect\citeauthoryear{{Sutherland} \& {Dopita}}{{Sutherland} \&
  {Dopita}}{1993}]{Sutherland:93}
{Sutherland} R.~S.,  {Dopita} M.~A.,  1993, \apjs, 88, 253

\bibitem[\protect\citeauthoryear{{Trenti}, {Stiavelli}, {Bouwens}, {Oesch},
  {Shull}, {Illingworth}, {Bradley} \& {Carollo}}{{Trenti}
  et~al.}{2010}]{Trenti.etal:10}
{Trenti} M.,  {Stiavelli} M.,  {Bouwens} R.~J.,  {Oesch} P.,  {Shull} J.~M.,
  {Illingworth} G.~D.,  {Bradley} L.~D.,    {Carollo} C.~M.,  2010, \apjl, 714,
  L202

\bibitem[\protect\citeauthoryear{{Wiersma}, {Schaye} \& {Smith}}{{Wiersma}
  et~al.}{2009}]{Wiersma.etal:09}
{Wiersma} R.~P.~C.,  {Schaye} J.,    {Smith} B.~D.,  2009, \mnras, 393, 99

\bibitem[\protect\citeauthoryear{{Wilkins}, {Bunker}, {Lorenzoni} \&
  {Caruana}}{{Wilkins} et~al.}{2011}]{Wilkins.etal:11a}
{Wilkins} S.~M.,  {Bunker} A.~J.,  {Lorenzoni} S.,    {Caruana} J.,  2011,
  \mnras, 411, 23

\bibitem[\protect\citeauthoryear{{Yajima}, {Choi} \& {Nagamine}}{{Yajima}
  et~al.}{2010}]{Yajima.etal:10}
{Yajima} H.,  {Choi} J.,    {Nagamine} K.,  2010, ArXiv e-prints

\bibitem[\protect\citeauthoryear{{Yan}, {Windhorst}, {Hathi}, {Cohen}, {Ryan},
  {O'Connell} \& {McCarthy}}{{Yan} et~al.}{2009}]{Yan&Windhorst.etal:09}
{Yan} H.,  {Windhorst} R.,  {Hathi} N.,  {Cohen} S.,  {Ryan} R.,  {O'Connell}
  R.,    {McCarthy} P.,  2009, ArXiv e-prints

\bibitem[\protect\citeauthoryear{{Yan} \& {Windhorst}}{{Yan} \&
  {Windhorst}}{2004}]{Yan&Windhorst:04}
{Yan} H.,  {Windhorst} R.~A.,  2004, \apjl, 612, L93

\end{thebibliography}

 \end{document}